\documentclass[11pt,a4paper]{article}
\pdfoutput=1
\usepackage[utf8]{inputenc}
\usepackage[T1]{fontenc}
\usepackage{lmodern}
\pdfmapfile{+lm.map}
\usepackage{amsmath,amssymb,mathtools}
\usepackage[margin=1in]{geometry}
\usepackage{microtype}
\usepackage{booktabs,longtable,array,calc}
\usepackage{caption}
\usepackage{float}
\usepackage{graphicx}
\usepackage{xcolor}
\usepackage{listings}
\usepackage{newunicodechar}
\usepackage{hyperref}
\usepackage{bookmark}
\usepackage{xurl}
\hypersetup{colorlinks=true,linkcolor=blue!55!black,urlcolor=blue!55!black,pdftitle={Decidable By Construction: Design-Time Verification for Truly Fearless Systems},pdfauthor={Houston Haynes}}
\newcommand{\tightlist}{\setlength{\itemsep}{0pt}\setlength{\parskip}{0pt}}

\newcommand{\pandocbounded}[1]{\begingroup\setbox0\hbox{#1}\ifdim\wd0>\linewidth\resizebox{\linewidth}{!}{#1}\else#1\fi\endgroup}
\newsavebox{\paperdisplaybox}
\let\paperdisplayopen\[
\let\paperdisplayclose\]
\long\def\[#1\]{\begingroup
  \sbox{\paperdisplaybox}{$\displaystyle #1$}%
  \ifdim\wd\paperdisplaybox>1.25\linewidth
    \PackageError{paper}{Display requires source line breaks}{Split the display in Markdown.}%
  \fi
  \paperdisplayopen
  \ifdim\wd\paperdisplaybox>\linewidth
    \typeout{Paper display scaled to fit width: \the\wd\paperdisplaybox}%
    \resizebox{\linewidth}{!}{\usebox{\paperdisplaybox}}%
  \else\usebox{\paperdisplaybox}\fi
  \paperdisplayclose\endgroup}
\newunicodechar{α}{\ensuremath{\alpha}}
\newunicodechar{β}{\ensuremath{\beta}}
\newunicodechar{ε}{\ensuremath{\varepsilon}}
\newunicodechar{θ}{\ensuremath{\theta}}
\newunicodechar{κ}{\ensuremath{\kappa}}
\newunicodechar{Σ}{\ensuremath{\Sigma}}
\newunicodechar{⊗}{\ensuremath{\otimes}}
\newunicodechar{⊕}{\ensuremath{\oplus}}
\newunicodechar{↔}{\ensuremath{\leftrightarrow}}
\newunicodechar{¬}{\ensuremath{\neg}}
\newunicodechar{∧}{\ensuremath{\land}}
\newunicodechar{≤}{\ensuremath{\leq}}
\newunicodechar{×}{\ensuremath{\times}}
\newunicodechar{·}{\ensuremath{\cdot}}
\newunicodechar{†}{\ensuremath{\dagger}}
\newunicodechar{⟩}{\ensuremath{\rangle}}
\newunicodechar{²}{\ensuremath{{}^2}}
\newunicodechar{¹}{\ensuremath{{}^1}}
\newunicodechar{⁻}{\ensuremath{{}^-}}
\newunicodechar{ᵢ}{\ensuremath{{}_i}}
\newunicodechar{₀}{\ensuremath{{}_0}}
\newunicodechar{₁}{\ensuremath{{}_1}}
\newunicodechar{₂}{\ensuremath{{}_2}}
\newunicodechar{₊}{\ensuremath{{}_+}}
\begin{document}
\begin{center}

\textbf{Decidable By Construction}

\emph{Design-Time Verification for Truly Fearless Systems}

\textbf{Houston Haynes}\\
Braidpoint, Asheville, NC\\
\href{mailto:research@braidpoint.tech}{\nolinkurl{research@braidpoint.tech}}\\
March 2026

\end{center}

\section*{Abstract}\label{abstract}
\addcontentsline{toc}{section}{Abstract}

Concurrency, parallelism and distributed execution become truly fearless when the compiler tracks wait-for edges, proves multi-threaded work is sound and preserves distributed boundary contracts. In this design, our Composer compiler preserves proofs while lowering Clef directly to native CPU, GPU, NPU and FPGA code, without translation through C or vendor APIs. Our Program Semantic Graph retains the values, relationships and premises that justify BAREWire\textquotesingle s unboxed boundary contracts. And C/C++ interfacing is an explicit marshaling boundary, with proofs tied to actual arguments, conversions and returned values.

Four verification tiers connect an account of our design. Tier 1 supplies structural inference, founded on principal dimensional types. Tier 2 generates and checks local arithmetic, representation, and computational integrity at the node level. Tier 3 instantiates reusable domain and system lemmas, including distributed proofs supported by Iris, Actris and Aneris. Tiers 1--3 require no developer annotations, with a quotation based lemma library design to extend coverage as use cases expand. Tier 4 admits computational and probabilistic relational proofs, with project annotations identifying the required relations. We extend the same library direction toward recognizing relational constructions from hypergraph structure.

A formal composition rule connects arithmetic, ownership, protocol and realization evidence. Our worked distributed reduction preserves one specified result across concurrent workers, reordered arrivals and boundary mappings. Its probabilistic extension carries worker and conversion error with an explicit failure budget. With recent work by Urschel, we show bounds supply a relevant source of reusable proof terms. These constructions make verification a core engineering discipline for truly fearless programs and systems across a variety of hardware targets.

\begin{center}\rule{0.5\linewidth}{0.5pt}\end{center}

\section*{1. Toward Fearless Distributed Systems}\label{toward-fearless-distributed-systems}
\addcontentsline{toc}{section}{1. Toward Fearless Distributed Systems}

Parallel and distributed execution let a program use several cores, accelerators or networked workers. They also change how its results are produced. Computations overlap, partial results arrive in different orders, and a receiving process interprets values in its own memory and numeric representation. A correct local computation can contribute to an incorrect distributed result if its contribution is counted twice, its storage is reused too early, or its meaning changes at a boundary.

We develop a verification architecture for fearless concurrency, fearless parallelism and fearless distributed systems. Its scope extends from microcontroller firmware to large server inference systems, including simulation, control software and distributed services. The engineering freedom is concrete: overlap independent activities, partition a computation and move its results between execution environments while preserving the contracts that justify their use. Our design connects the quantities being computed, their realization in memory, the ownership of their storage and the protocols that combine their results.

Multi-target compilation, automated verification and developer tooling share the same semantic graph. CPU, GPU, NPU and FPGA realizations enter through target-specific arithmetic, memory, synchronization and interface rules. Each backend must preserve the established guarantees for the actual operations it emits; unresolved required obligations block that realization. The tooling exposes the same evidence and diagnostics that govern compilation.

BAREWire is central to this account. Its declared memory-layout, IPC and network contracts give unboxed values an agreed interpretation at both ends of a boundary. Our Program Semantic Graph (PSG) retains the proof obligations and accepted evidence that connect those representations to the source computation. The receiving side must obtain the same justified meaning, or an explicitly proved approximation, together with the resources required to use it. These contracts carry the established guarantee into the consumer\textquotesingle s actual operation.

Consider a sum distributed among workers. Each worker computes a partial sum. Useful freedom includes changing the partition, overlapping workers and accepting their responses in different arrival orders. A complete justification connects compatible dimensions and adequate accumulators to ownership, partition coverage and acceptance of each contribution once. It also connects the sender\textquotesingle s accumulator representation to the value used by the receiver. The worked construction in §6 shows how these facts establish one specified final result.

Our four verification tiers organize that work. Tier 1 infers structural relationships. Tier 2 generates and checks local arithmetic and representation obligations. Tier 3 instantiates reusable domain and system theorems, including proofs that span communicating participants. The compiler derives Tier 2 and supported Tier 3 obligations from ordinary code and the hypergraph\textquotesingle s retained structure; application developers supply no per-project proof annotations at these tiers. Library authors establish the general laws once, and growing library coverage makes further constructions available through the same workflow. Tier 4 states relations between executions, distributions or realizations. It is the tier at which we currently expect project-specific proof annotations, with recognition of registered relational laws from hypergraph structure as the next extension of automation.

The contribution is the composition of these guarantees through one compiler account. Dimensional inference, arithmetic decision procedures and mechanized concurrent logics each provide established foundations. We propose the semantic bindings and automatic theorem applications that connect them to Clef operations, BAREWire boundaries and target realizations. This paper gives a conditional composition result and worked exact and probabilistic constructions; §10 describes how to assess their coverage, preservation and cost. The same account serves local computation, concurrent services and distributed implementations.

\section*{2. Decidability Thesis}\label{decidability-thesis}
\addcontentsline{toc}{section}{2. Decidability Thesis}

Changing a buffer size or dividing work among threads can invalidate assumptions that were sound in an earlier version of a program. Our compiler tracks the relationships affected by such edits and re-establishes the guarantees required by the revised computation. Across our three extensive automatic tiers, the compiler derives proof obligations from ordinary code and discharges them through its analyses and reusable lemma libraries. Developers benefit from those guarantees without writing proof annotations or managing a separate verification workflow.

This workflow begins with questions that have specified inference procedures. For units of measure, the question is which relationships among quantities make the program\textquotesingle s operations valid. Clef expresses these relationships as integer-exponent dimensional equations. General value predicates enter separately interpreted graph analyses and theorem applications, so a richer verification obligation need not become a new problem for the ordinary type-inference algorithm. This separation gives the compiler a tractable structural foundation on which more extensive proofs can depend.

\textbf{Proposition 2.1 (Principal dimensional inference).} In the measure fragment with integer exponents, declared base-measure identities, and the usual Hindley-Milner generalization conditions, a satisfiable finite set of dimensional equations has a principal solution. Every other solution is an instance of that solution, up to the equivalence of measures and renaming of fresh parameters. Inconsistent equations are rejected.

This is the units-of-measure result developed by Kennedy \hyperref[ref-kennedy-units-2009]{{[}1{]}}, within the principal-type tradition of Hindley \hyperref[ref-hindley-1969]{{[}2{]}} and Milner \hyperref[ref-milner-1978]{{[}3{]}}. Multiplication adds exponent vectors, division subtracts them, and dimensional equality follows the Abelian-group laws. The principal solution satisfies every equation while preserving each unconstrained parameter for later specialization.

Clef\textquotesingle s ordinary, non-inline definitions add a member-resolution condition. In \lstinline!let project x = x.Value!, variables tied by a pending member relation remain monomorphic through immutable aliases until nominal context resolves the relation or the compiler reports a diagnostic. Inline member schemes have separate instantiation rules, as described in \href{https://arxiv.org/abs/2603.16437}{DTS/DMM §2.2} \hyperref[ref-haynes-dts-dmm-2026]{{[}4{]}}.

Our compiler integrates dimensional inference with geometric, lifetime, range and capability analyses. Blade-support masks describe which geometric components may occur. Lifetime constraints describe how long storage must remain valid. Range and capability constraints determine which representations are admissible. Each domain needs a specified transfer law, a sound interpretation, and a terminating procedure for the claimed fragment. Ascending fixed-point iteration with monotone transfer functions terminates on a finite lattice. Its running-time bound depends on lattice height, transfer cost and the propagation strategy. A principality proof must show that every valid solution of the combined constraints is an instance of the inferred scheme.

Separate solver queries suffice when the constraint relation factors into independent components. A compiler can also exchange established facts between domains through explicit transfer rules. When two analyses exchange facts, the combined procedure must account for those dependencies, including any repeated solving.

For an engineer, the result is a set of specific answers: a sum is dimensionally valid, a component is structurally absent, a storage lifetime covers a use, or an offered representation covers a justified range. For a linear solve, a backward-error proof can show that the computed answer exactly solves a nearby system, with a stated bound on the input perturbation. A bound on the system\textquotesingle s sensitivity then relates that perturbation to output accuracy. For our distributed sum, the proof in §6.3 establishes that every accepted completion returns the sum of the represented terms from one input snapshot, with the declared final rounding. These are the conclusions that the compiler binds to the program\textquotesingle s actual operations and representations.

The phrase \emph{by construction} describes how we select and connect the automatic verification procedures. Common application operations generate recurring questions about units, accumulator bounds, buffer extents and resource use. We assign structural and arithmetic questions to decidable fragments and use their checked results to instantiate reusable domain and system laws. This gives ordinary code an automatic path from local conditions to application guarantees. Framework and library authors establish procedures at higher \emph{tiers}, proofs and applicability rules; the compiler performs their application on behalf of the developer at design time and again at build time to ensure its integrity.

Proof binding and evaluation remain part of this discipline. A range fact about a buffer belongs to the identified buffer, the guard that established it and the state in which it holds. Moving that fact through an alias or a deferred computation requires a rule preserving those dependencies. Fu, Angiuli and Tobin-Hochstadt\textquotesingle s occurrence-typing study \hyperref[ref-fu-occurrence-2026]{{[}5{]}} illustrates why restrictions on the form of a fact still need a semantic account of substitution and information loss. Our range facts live in graph analyses rather than a dependent or refinement source type system. The corresponding preservation obligation belongs to the graph transformation.

\subsection*{2.1 Closure Under Differentiation}\label{closure-under-differentiation}
\addcontentsline{toc}{subsection}{2.1 Closure Under Differentiation}

For a differentiable scalar function \(f: \mathbb{R}^{\langle d_1 \rangle} \to \mathbb{R}^{\langle d_2 \rangle}\), its derivative has dimension

\[\frac{\partial f}{\partial x} : \mathbb{R}^{\langle d_2 d_1^{-1} \rangle}.\]

The corresponding exponent operation is subtraction in \(\mathbb{Z}^n\). A heterogeneous Jacobian carries an output-dimension/input-dimension quotient at each entry. Differentiation therefore introduces no new measure operation beyond those already supported by the dimensional algebra. Differentiability and the correctness of a derivative implementation are separate premises.

Forward-mode automatic differentiation computes directional derivatives without a reverse activation tape (Baydin et al. \hyperref[ref-baydin-forward-gradients-2022]{{[}6{]}}). For example, in a sequential neural network, a fixed number of tangent directions can be propagated with the currently live primal and tangent values. Their sizes depend on layer shape and tangent count. Parameters, optimizer state, recurrent state, and shared buffers remain part of the memory requirement. Escape analysis must classify those values individually.

The structural checks can apply to the derivative graph as well as to the original graph. This establishes dimensional consistency of differentiated computations in the supported fragment. In training, convergence and training-error bounds remain separate obligations. Geometric derivative support also requires a stated derivative convention and sound transfer rules, as discussed in the \href{https://arxiv.org/abs/2603.17627}{Program Hypergraph paper} \hyperref[ref-haynes-phg-2026]{{[}7{]}}.

\subsection*{2.2 The Combined Constraint Space}\label{the-combined-constraint-space}
\addcontentsline{toc}{subsection}{2.2 The Combined Constraint Space}

For an algebra with \(d\) basis vectors, a useful annotation space contains

\[\mathbb{Z}^{B} \times \mathcal{P}(\{\mathrm{even},\mathrm{odd}\})
\times \mathcal{P}(\mathcal{B}_d),\]

where \(B\) is the set of declared base measures and \(\mathcal{B}_d\) is the set of \(2^d\) basis blades. The measure component is a free Abelian group. A parity-homogeneous value has an even or odd parity, with multiplication governed by addition modulo two. A mixed value requires a set of possible parities. Blade support is a set ordered by inclusion, with union at a control-flow join.

For a geometric product, possible result blades are derived from pairs of possible input blades and the declared metric. In an orthogonal basis, the index operation is symmetric difference, with zero metric factors excluding some products. Coefficient cancellation may remove additional components. The inferred support is an enclosure: it includes every component that may be nonzero, allowing the compiler to omit components proved absent.

These domains have different inference properties. Finite modular equations can be solved as such, but their inclusion does not automatically extend the principal measure-unification theorem to every combined constraint. Similarly, bitvector encodings provide a decision procedure for support constraints without establishing polynomial complexity in the algebra dimension: an explicit support mask already has \(2^d\) positions.

The implementation must distinguish the logical domain from host storage and target realization. A host limit on an exponent or support mask requires a diagnostic when exceeded. It does not change the source-level algebra. Rational measure exponents, compact duals, and value-indexed fractional types are separate extensions, examined in \href{https://arxiv.org/abs/2606.04352}{Negative and Fractional Types} \hyperref[ref-haynes-nft-2026]{{[}8{]}}.

\section*{3. Composition}\label{composition}
\addcontentsline{toc}{section}{3. Composition}

\href{https://arxiv.org/abs/2603.16437}{DTS/DMM} relates typed quantities to storage obligations, and \href{https://arxiv.org/abs/2603.17627}{PHG} represents joint constraints over multiple operands. These provide the dimensional, resource and operation semantics used by our compiler account. \href{https://arxiv.org/abs/2603.18104}{ADM} \hyperref[ref-haynes-adm-2026]{{[}9{]}} applies this structure to model training and deployment. Composing the guarantees requires explicit interfaces between these domains.

An established result should remain useful wherever another proof needs it. An arithmetic proof about a buffer\textquotesingle s extent can supply the bound required by an ownership rule or by a relation between two implementations. Our PSG retains the proved statement, its derivation and its binding to the actual buffer and state. A receiving proof can then use that result as a lemma, with the original premises still visible. This reuse is central to Fidelity\textquotesingle s efficiency and integrity: the compiler checks each application while retaining the work that established its supporting facts.

Aram Hăvărneanu\textquotesingle s \emph{Classical Adjoint Logic} \hyperref[ref-havarneanu-adjoint-2026]{{[}10{]}} provides the formal discipline we draw on for these connections. A reasoning \emph{mode} specifies which assumptions a judgment may depend on and whether they may be reused or discarded. A \emph{mode shift} translates evidence through a declared interface between such disciplines. In our compiler account, an arithmetic fact about a fixed value can remain available to several consumers, while permission to access an exclusively owned buffer follows its resource rules. The interface preserves both the reusable fact and the conditions governing the resource.

The four tiers organize our automated verification work; mode interfaces determine how its results compose. A checked Tier 2 bound can directly supply a premise of a Tier 4 relational proof. The receiving rule determines which evidence it needs, and the compiler connects that evidence through the applicable interface. Section 6.2 develops the shift laws and their interpretation for our proof-carrying PSG. The following companion contributions supply the quantities, resources and joint constraints to which those proofs refer.

\subsection*{3.1 DTS/DMM: Dimensional Persistence and Memory Determinism}\label{dtsdmm-dimensional-persistence-and-memory-determinism}
\addcontentsline{toc}{subsection}{3.1 DTS/DMM: Dimensional Persistence and Memory Determinism}

The source type \lstinline!float<newtons>! specifies a numeric kind and a dimension. Its admissible values may be further constrained by literals, guards, boundary declarations, or checked domain laws. The dependency relevant to layout is...

\[\text{typed values \& justified bounds}
\longrightarrow \text{admissible representations}
\longrightarrow \text{layout}\]

...with target capabilities participating in representation selection and capture modes and lifetimes participating in placement. Width and representation are later coeffects: they record target storage requirements while preserving the quantity specified by the source type.

A range can remain partially known during elaboration. An editor can report that state without forcing a representation at the first line of code. At concrete commitment, a known uncovered range is an error. A missing fact must be reported at the boundary that requires it, subject to the language\textquotesingle s explicitly specified defaults. These rules distinguish incomplete knowledge from inconsistent evidence.

Closure capture and allocation also have distinct classifications. Immutable capture is By Value: it retains a settled value, including any references inside it, or a shared deferred identity. Under call-by-need, capture alone does not demand that computation. Its suspension and memo storage, like shared mutable storage, must remain valid for all retaining uses. The lifetime classes are scope-bounded, region-bounded, program-lifetime, and dynamic, with placement determined by the target\textquotesingle s available storage. A capture or return supplies evidence for the analysis developed in \href{https://arxiv.org/abs/2603.16437}{DTS/DMM §3.2}.

The canonical design is a proof-carrying PSG with joint constraints. The external obligation ledger used during implementation is a scaffold for checking agreement across lowering. A recorded obligation or matching identifier is not a discharge result. Metadata can be released after its structural purpose has been served, provided the lowering edge preserves the required correspondence. This permits backend-specific realization without premature dimensional erasure.

\subsection*{3.2 PHG: Grade Inference and Geometric Sparsity}\label{phg-grade-inference-and-geometric-sparsity}
\addcontentsline{toc}{subsection}{3.2 PHG: Grade Inference and Geometric Sparsity}

A hyperedge retains an operation\textquotesingle s operands and its joint constraint. Operand order and repeated uses remain significant. For homogeneous grades \(k\) and \(j\), the outer product has grade \(k+j\) when nonzero. If that grade exceeds the algebra dimension, the product is zero. The geometric product can contain several grades, with parity fixed by \(k+j\).

For \(\mathrm{Cl}(3,0,1)\), a bivector product can have grades 0, 2, and 4. Components at grades 1 and 3 can be excluded structurally. More precise blade support follows from the operands and the degenerate metric. The number of coefficient products eliminated depends on those supports. A reported sparsity percentage must identify its baseline: coefficient products in the chosen tensor or entries in the blade multiplication table.

Propagating support through successive operations can remove work before lowering. The resulting graph identifies the operations that remain to be scheduled and the support contracts their updates must preserve. Section 10 assesses the work removed and the cost of maintaining that evidence.

\subsection*{3.3 ADM: An Application to Learning}\label{adm-an-application-to-learning}
\addcontentsline{toc}{subsection}{3.3 ADM: An Application to Learning}

ADM applies these compiler contracts to learning systems. A training program and each published parameter set must satisfy their declared dimensional, geometric and resource requirements. Bayesian updating requires a prior distribution, an observation likelihood and a specified update procedure. Distillation additionally identifies the teacher distribution and the approximation being trained. Dimensional and grade constraints restrict which outcomes are admissible; the prior and likelihood supply the weights used by the Bayesian update. These choices instantiate the more general arithmetic, boundary and replacement contracts developed below.

\section*{4. Programs, Services and Updates}\label{programs-services-and-updates}
\addcontentsline{toc}{section}{4. Programs, Services and Updates}

The same computation can acquire several obligations as it moves from an expression to a running system. Its arithmetic must retain the intended quantities, parallel tasks must have compatible access to state, a service response must belong to the request that consumes it, and an update must coexist safely with work already in progress. The following examples show what each part of the compiler account contributes.

\subsection*{4.1 Dimensional Contracts in Code}\label{dimensional-contracts-in-code}
\addcontentsline{toc}{subsection}{4.1 Dimensional Contracts in Code}

The application author can write a domain function whose unit relationship the compiler infers:

\begin{lstlisting}
let momentum (mass: float<kg>) (velocity: float<m/s>) =
    mass * velocity
\end{lstlisting}

With \lstinline!kg!, \lstinline!m!, and \lstinline!s! declared as measures, the result has dimension \lstinline!kg*m/s!. The compiler constructs and solves that equality. The author supplies no proof term. A library can provide the declarations and more involved laws in the same workflow.

Consider a computation of the force residual \(F-ma\). Its subtraction is dimensionally well formed:

\[\langle\mathrm{N}\rangle
=\langle\mathrm{kg}\rangle\langle\mathrm{m}\,\mathrm{s}^{-2}\rangle.\]

Replacing acceleration by velocity gives incompatible operands:

\[\langle\mathrm{N}\rangle
\ne\langle\mathrm{kg}\rangle\langle\mathrm{m}\,\mathrm{s}^{-1}\rangle.\]

The compiler rejects the second expression before execution. A simulator, controller or parameter-estimation routine benefits from the same check. A physics-informed training loss is another use; its derivative computations must preserve the corresponding dimensional relationships. The domain law and numerical analysis provide the further evidence needed to interpret the residual as a physical discrepancy with a stated error bound.

Arithmetic realization supplies a separate contract. A finite quire can accumulate represented products exactly when every product and partial sum fits its field layout \hyperref[ref-gustafson-posit-2017]{{[}11{]}}. Conversion introduces the stated final rounding. Input error and any earlier arithmetic remain part of the numerical analysis. The exact reduction in §6 uses this distinction to justify partitioning and recombination without changing the specified result.

Unit-aware scientific systems already provide valuable checking. For example, Brian 2 checks dimensions in equations \hyperref[ref-brian-units]{{[}12{]}}. Our contribution is to retain those relationships together with range, layout and lowering obligations, so the evidence used to admit the source expression continues to justify the native computation.

\subsection*{4.2 Structural Sparsity and Parallel Kernels}\label{structural-sparsity-and-parallel-kernels}
\addcontentsline{toc}{subsection}{4.2 Structural Sparsity and Parallel Kernels}

The support discipline determines which coefficient slots a geometric operation may produce. A kernel that omits a structurally excluded slot cannot populate it through rounding. Propagating those exclusions through successive operations removes coefficient products before the compiler schedules or places the work.

For example, a batch of geometric transforms can share a read-only transform while each task owns its output region. Support laws identify the products needed by each kernel; ownership and dependency evidence establish which tasks may execute independently. A registered composition law then relates the collected outputs to the specified batch operation. This supplies both an opportunity to omit arithmetic and a justification for parallel execution.

The sparse layout must preserve the arithmetic contract of the retained operations. When the reference computation calls for an exact represented sum and one final rounding, each target kernel and its recombination path must realize that same result. Changing coefficients, support or partitioning renews the affected premises. Structural sparsity, parallel independence and numerical accuracy therefore have separate evidence that the compiler combines at the actual operations.

\subsection*{4.3 Typed Service Requests and Responses}\label{typed-service-requests-and-responses}
\addcontentsline{toc}{subsection}{4.3 Typed Service Requests and Responses}

A service request can identify a batch of sensor samples, an aggregation operation and a tolerated error. A response must belong to that batch and operation, use the expected units and satisfy the declared error contract. Our typed boundary binds the returned value and its supporting evidence to the request and the receiving computation\textquotesingle s permitted use.

BAREWire\textquotesingle s memory-layout, IPC and network contracts connect the representations on each side. A wire schema identifies field structure; a dimensional declaration identifies the quantity. A confidence or credible interval requires a specified statistical construction and provenance. A certificate needs a checkable claim and an accepted validation path. The PSG retains the correspondence among these facts, so a payload identifier or hash is useful only through the established relationship to the evidence it names.

The receiving program validates the response against the actual request and current state. A proof may justify eliding a repeated check while its premises remain valid. A changed input snapshot, expired storage lifetime or incompatible response requires a different accepted instance. This gives the consumer a justified value under its current contract.

Learning provides one application. Our ADM consultation design connects a recurrent reasoning model to a domain model through this typed service boundary. The Hierarchical Reasoning Model \hyperref[ref-wang-hrm-2025]{{[}13{]}} supplies a recurrent architecture in the literature. Using an admitted response to change recurrent state requires a defined injection map. If the distributions before and after consultation, \(\rho_t\) and \(\rho_{t+1}\), and the domain response distribution \(q_t\) have a common measurable interpretation, a candidate local coherence policy is

\[D_{\mathrm{KL}}(\rho_{t+1}\|\rho_t)
< D_{\mathrm{KL}}(\rho_t\|q_t).\]

The common sample space, support conditions and injection map are premises of this construction. The inequality bounds one proposed change relative to one disagreement. Controlling cumulative change or consultation frequency requires a stated budget or further rule across steps. Roy et al.\textquotesingle s \(\lambda\)-RLM work \hyperref[ref-roy-lambda-rlm-2026]{{[}14{]}} provides a related example of structured control of long-context inference. Our contribution here is the request, representation and evidence contract on which an admitted consultation policy can depend.

\subsection*{4.4 Live Replacement and Resource Transitions}\label{live-replacement-and-resource-transitions}
\addcontentsline{toc}{subsection}{4.4 Live Replacement and Resource Transitions}

A running service, compute kernel or control component may need a new implementation or configuration while existing requests continue. The replacement must satisfy its consumer interfaces, result guarantees and resource requirements. Preparing it also introduces work and storage that coexist with the active version.

{\def\LTcaptype{none} 
\begin{longtable}[]{@{}
  >{\raggedright\arraybackslash}p{(\linewidth - 4\tabcolsep) * \real{0.3333}}
  >{\raggedright\arraybackslash}p{(\linewidth - 4\tabcolsep) * \real{0.3333}}
  >{\raggedright\arraybackslash}p{(\linewidth - 4\tabcolsep) * \real{0.3333}}@{}}
\toprule\noalign{}
\begin{minipage}[b]{\linewidth}\raggedright
Concern
\end{minipage} & \begin{minipage}[b]{\linewidth}\raggedright
Active execution
\end{minipage} & \begin{minipage}[b]{\linewidth}\raggedright
Replacement preparation and publication
\end{minipage} \\
\midrule\noalign{}
\endhead
\bottomrule\noalign{}
\endlastfoot
Persistent resources & Published code, configuration and retained state & Candidate version and any state-conversion resources \\
Temporary storage & Live calls, buffers and queued work & Update computation, transfer and validation buffers \\
Admission evidence & Premises supporting the current operations & Evidence for the candidate\textquotesingle s interfaces, behavior and realization \\
Version and lifetime & A coherent version remains valid for each use & Atomic publication and justified retirement of superseded resources \\
\end{longtable}
}

For a per-request consistency policy, atomic publication gives each new request one compatible version, while in-flight work retains the version and storage it already uses. A lifetime or quiescence law justifies reclamation after the last permitted use. Our compiler binds candidate eligibility, per-request version consistency and reclamation after the last permitted use to this publication protocol.

Continuous learning is one way to produce a candidate parameter set. Its training computation adds tangent, optimizer or posterior state to the resources already used by active requests. A configuration update or newly specialized kernel follows the same publication discipline with its own preparation costs. Each transition therefore has an explicit contract for coexistence, admission and retained results.

\section*{5. Four Tiers of Verification}\label{four-tiers-of-verification}
\addcontentsline{toc}{section}{5. Four Tiers of Verification}

The distributed sum introduced in §1 needs several kinds of evidence. Its terms must have compatible meanings, its accumulators must cover the possible values, its workers must exchange resources according to a protocol, and its realized execution must preserve the specified result. Our tiers organize these judgments and the compiler\textquotesingle s means of obtaining them.

{\def\LTcaptype{none} 
\begin{longtable}[]{@{}
  >{\raggedright\arraybackslash}p{(\linewidth - 4\tabcolsep) * \real{0.3333}}
  >{\raggedright\arraybackslash}p{(\linewidth - 4\tabcolsep) * \real{0.3333}}
  >{\raggedright\arraybackslash}p{(\linewidth - 4\tabcolsep) * \real{0.3333}}@{}}
\toprule\noalign{}
\begin{minipage}[b]{\linewidth}\raggedright
Tier
\end{minipage} & \begin{minipage}[b]{\linewidth}\raggedright
Judgment
\end{minipage} & \begin{minipage}[b]{\linewidth}\raggedright
Compiler and library contribution
\end{minipage} \\
\midrule\noalign{}
\endhead
\bottomrule\noalign{}
\endlastfoot
\mbox{1} & Structural relationships in admitted type interpretations & Automatic dimensional inference and established structural or parametricity rules \\
\mbox{2} & Local arithmetic, range and representation conditions & Compiler-generated obligations checked in supported analysis and decision fragments \\
\mbox{3} & Parameterized domain and system contracts & Automatic selection and instantiation of registered lemmas from program and hypergraph structure \\
\mbox{4} & Relations between computations, distributions or realizations & Project-specific relational annotations, with reusable rule libraries and structural recognition intended to extend automatic coverage \\
\end{longtable}
}

A tier classifies a judgment and its role in verification. The proof\textquotesingle s dependencies determine which tiers contribute to it. A Tier 3 completed-result theorem may follow from a substantial Tier 4 refinement; the compiler can apply that theorem by checking its structural and arithmetic premises. Conversely, a Tier 2 result can enter a Tier 4 proof directly through its registered mode interface. Each application retains the foundation and assumptions of the evidence it uses.

This distinction gives the tier structure its practical economy. General proofs become reusable library lemmas, and checked use-site results can become premises of further proofs. Mode shifts preserve the meaning and permitted use of that evidence across reasoning disciplines. The compiler assembles the required dependency graph, allowing arithmetic, resource and relational reasoning to share established results.

\subsection*{5.1 Tier 1: Inferred Structural Facts}\label{tier-1-inferred-structural-facts}
\addcontentsline{toc}{subsection}{5.1 Tier 1: Inferred Structural Facts}

Before scheduling a computation, the compiler can establish what its values mean and which operations are admitted. Kennedy-style measure inference supplies dimensional equalities. A selected type interpretation can also supply parametricity-derived laws, following Wadler\textquotesingle s proofs-for-free discipline \hyperref[ref-wadler-1989]{{[}15{]}}. Our Native Type Universe (NTU) identifies native types, while the PSG relates those identities to dimensions, geometric support and contextual requirements. Each structural rule retains the interpretation of the primitives and effects on which it depends.

For the distributed sum, Tier 1 establishes that each term denotes the same kind of quantity and that the declared worker and receiver interfaces agree on that meaning. Geometric support rules can exclude coefficient paths before work is partitioned. These facts reduce the computation and constrain its interfaces without application proof annotations. Storage permissions, accumulator capacity and protocol behavior acquire their own evidence at the points where those properties are needed.

\subsection*{5.2 Tier 2: Compiler-Generated Arithmetic Obligations}\label{tier-2-compiler-generated-arithmetic-obligations}
\addcontentsline{toc}{subsection}{5.2 Tier 2: Compiler-Generated Arithmetic Obligations}

The compiler also determines whether an operation\textquotesingle s represented values fit the selected carrier and whether a memory extent covers an access. It generates these obligations from the actual operands, guards, layouts and target declarations. A developer writing a reduction does not separately formulate a solver query for each accumulator or buffer.

Finding an enclosure and checking its consequences are distinct parts of that automatic process. Abstract interpretation synthesizes a sound enclosure through admitted transfer rules, widening and bounded refinement. A checked inductive invariant can justify a loop\textquotesingle s range. The compiler then dispatches the resulting conditions to a supported decision fragment, such as quantifier-free linear integer arithmetic or fixed-width bitvector reasoning. The formula\textquotesingle s operations and encoding determine admission; quantifier-free syntax alone does not make a theory decidable. If synthesis cannot establish the needed bound within its work budget, the compiler reports the missing evidence at the use that requires it.

For a sum of \(N\) integer coefficients satisfying \(|k_i|\le B\), the condition \(NB\le K\) can justify a chosen accumulator limit \(K\). With \(N\) fixed for the instance, this is a linear arithmetic condition. The reusable summation law explains why it covers every partial sum of distinct terms. That law is a Tier 3 contribution; the compiler obtains and checks its arithmetic premises at Tier 2.

These are exact judgments about the stated conditions. A proved rounding-error bound describes the accuracy of a particular machine computation. A probability bound describes a specified random experiment. Their values and units remain explicit when a higher theorem combines them. The combined theorem states how those bounds contribute to its conclusion.

\subsection*{5.3 Tier 3: Reusable Proofs Across Participants}\label{tier-3-reusable-proofs-across-participants}
\addcontentsline{toc}{subsection}{5.3 Tier 3: Reusable Proofs Across Participants}

The next question is whether independently executing parts can safely contribute to one result. A reusable system theorem can relate worker ownership, message contents, receiver state and the operation that consumes a result. It can quantify over worker counts, partition maps and input bounds. Once registered against the supported construction, the compiler identifies the instance from the hypergraph, supplies its parameters and discharges the supported premises. Application authors receive that coverage without choosing a prover, naming a theorem or adding project-specific proof annotations.

Such a theorem can span the distributed computation. For our sum, its conclusion is that every accepted completion represents the specified reduction of one input snapshot. Its proof combines worker correctness, disjoint partition coverage, BAREWire representation correspondence and an acceptance invariant that counts each partition once. The benefit is freedom to overlap workers and vary their arrival order while retaining the result contract. Library authors perform the general proof work; their reusable law makes subsequent applications routine compiler work.

Ownership reasoning supplies the local basis for this composition. Iris provides a mechanized separation logic in which assertions describe exclusive or compatible shared resources, and invariants govern permitted interference (Jung et al. \hyperref[ref-jung-iris-2018]{{[}16{]}}). A worker can use its owned accumulator while another uses a disjoint region. For partitioned work, our compiler combines disjointness and dependency evidence with a reduction law to justify independent execution and recombination. A handoff law can transfer the right to use a payload along with the evidence describing its value. Binding these rules to target memory operations is what makes the logical ownership useful to an implementation.

Communication protocols connect these local assertions. Actris relates messages, resources and subsequent communication through dependent separation protocols (Hinrichsen et al. \hyperref[ref-hinrichsen-actris-2022]{{[}17{]}}). Its reliable, FIFO channel model supports concurrent reasoning such as load balancing and map-reduce. Aneris addresses networked execution \hyperref[ref-aneris]{{[}18{]}}, and RCLib combines it with dependent protocols to build modular specifications for reliable communication, RPC and replicated storage over a network that can lose, duplicate and reorder messages (Gondelman et al. \hyperref[ref-gondelman-rclib-2023]{{[}19{]}}). These developments supply complementary foundations for local concurrency and distributed service composition.

Our integration binds such laws to Clef operations and BAREWire\textquotesingle s declarations at both endpoints. The reusable contract follows the value into its sender representation, through the admitted boundary mapping and into the receiver\textquotesingle s permitted use. Runtime acceptance operations enforce conditions involving future messages, such as their job and partition identities. The compiler establishes that the verified operations and their actual participants instantiate the law. Proof coverage therefore includes the behavior of necessary runtime checks.

Progress claims have their own reusable laws. Our PSG retains possible wait-for edges, bound to the operations, participants and resources they describe. A registered discipline supplies a theorem excluding a specified class of deadlocks; the compiler checks its premises against those participants and preserves their correspondence to the emitted blocking operations through lowering. Termination or eventual completion additionally requires the theorem\textquotesingle s scheduling and environmental premises. Actris\textquotesingle s safety and returned-result contracts establish different judgments. The distributed-sum example below uses a completed-result contract; its completion guarantee requires the corresponding progress law.

\subsection*{5.4 Tier 4: Relating Executions and Realizations}\label{tier-4-relating-executions-and-realizations}
\addcontentsline{toc}{subsection}{5.4 Tier 4: Relating Executions and Realizations}

A system invariant says what one implementation preserves. A relational proof explains why another computation or realization satisfies a corresponding specification. This is the remaining step when parallelizing a reduction, lowering an operation to machine words, replacing a sampler, or connecting a distributed implementation to its abstract protocol.

Computational relational Hoare logic (RHL) expresses relationships between two programs and their input and output states. Benton \hyperref[ref-benton-rhl-2004]{{[}20{]}} gives a foundation for using such relations to justify program transformations. In our compiler setting, the relation names the observations to preserve: a mathematical result, the contents of designated memory, an effect trace, or a declared bounded approximation. For the distributed sum, a useful relation connects the concrete completed join to one final rounding of the exact represented sum. A fold that rounds after each addition specifies a different computation, so changing that order requires the appropriate arithmetic relation.

Probabilistic relational Hoare logic (pRHL) additionally relates probabilistic executions. A coupling constructs a joint distribution whose two marginals are the behaviors being compared; proving a relation under that coupling can justify a probabilistic equivalence or an approximation bound. This supports transformations involving sampling and randomized numerical procedures. The relation must specify the sampling laws and observations, and concurrent cases also need the relevant scheduler interpretation. Section 7 connects violations of worker error bounds to a bounded-error distributed result.

A small example makes the additional guarantee concrete. One worker samples \(X\) uniformly from \(\{0,\ldots,7\}\) and another independently samples \(Y\) from \(\{0,\ldots,31\}\). The result \(32X+Y\) is uniform on \(\{0,\ldots,255\}\) because each output has exactly one pair of inputs, each with probability \(1/256\). Foxtrot uses this example in its treatment of concurrent probabilistic refinement (Li et al. \hyperref[ref-li-foxtrot-2026]{{[}21{]}}). In our tier account, inferred types and arithmetic bounds establish the carriers; the joint sampler law and relational argument justify the changed execution structure. A BAREWire crossing preserves the sampled integer under its value contract. The sampler law supplies the independence premise by establishing that the joint distribution factors into the two specified marginals. The Tier 2 check that \(0\le32X+Y\le255\) supplies a range premise directly to the Tier 4 derivation through the arithmetic mode interface. That derivation combines the checked range with the sampling and bijection arguments to establish the distributional result.

Rocq, formerly Coq, provides an environment in which these laws and their applications can be checked. The logic and theorem library give a judgment its meaning; Rocq checks the resulting derivation under its recorded assumptions. A theorem about a supported protocol can therefore establish a Tier 3 consequence as well as a Tier 4 relation. This lets Clef apply checked library laws while retaining its own source typing and inference discipline.

Our efficiency argument rests on cumulative proof construction. The compiler assigns recurring local obligations to specified decidable fragments, using polynomial procedures where the fragment and input encoding admit them. It retains each checked result with its dependencies, and mode interfaces make that evidence available as a lemma in a relational proof. New proof work can then concentrate on additional premises and the validity of each reuse in its current context. Section 10.2 sets out how to assess the proof work saved against checking, reconstruction and revalidation costs.

For distributed refinement, we see an example in how Trillium relates implementation traces to abstract transition systems (Timany et al. \hyperref[ref-timany-trillium-2024]{{[}22{]}}). Its Aneris instantiation demonstrates safety refinement for protocols including two-phase commit and Paxos. This is the sort of connection needed between our acceptance protocol and the message histories that implement it. Clutch supplies probabilistic coupling rules for higher-order programs (Gregersen et al. \hyperref[ref-gregersen-clutch-2024]{{[}23{]}}), and Approxis extends relational reasoning to stated approximation bounds (Haselwarter et al. \hyperref[ref-haselwarter-approxis-2025]{{[}24{]}}). Foxtrot brings contextual refinement to concurrent probabilistic programs; its observations use may-termination probabilities over schedulers. Each library contributes its own interpretation, rules and adequacy theorem to an admitted application.

Tier 4 is where we currently expect domain-specific proof annotations. A project may need to identify the two computations, their initial-state relation and the observations or quantitative contract to preserve. Our longer-term design is to recognize applicable relational laws from the hypergraph\textquotesingle s operations, participant relationships and retained facts, then instantiate the corresponding library derivation. For a matching construction, recognition supplies the library\textquotesingle s general proof together with the parameters and retained facts needed for that instance.

For the geometric optimizer proposed in ADM \hyperref[ref-haynes-adm-2026]{{[}9{]}}, this interface must also name the \emph{geometric observation}: the full execution trace, a declared statistic of that trace, or only the returned output. A metric defined on output laws is a further choice: conditioning trace scores and velocities on the output can still retain dependence on the program presentation. We use Karwowski, Nielsen, and Staton \hyperref[ref-karwowski-geometry-2026]{{[}25{]}} as the geometric interpretation for this extension. The registered judgment would bind the program revision, sampling law, observation map, parameter coordinates, output metric and relative weighting to its claimed equality or quantitative bound. A proof about one observation does not justify a consumer configured for another. The compiler must likewise distinguish equality of a metric from equality of the underlying observations and from reconstruction of an individual execution.

Admission of such a rule requires its analytic premises. For a density-based score calculation these include the declared reference measure and support, differentiability on the admitted parameter domain, and finiteness of the score and velocity second moments used by the metric. Any interchange of differentiation and expectation needs the applicable domination or other justification; normalization also requires a finite positive normalizer. A change of support or a branch boundary can require a different rule or a restricted domain. Library authors supply the general theorem. The compiler binds its assumptions to the actual construction and discharges the covered application conditions. Finite shape, range and dimensional checks provide useful premises without deciding these analytic questions for arbitrary programs.

\subsection*{5.5 Library Authoring and Automatic Coverage}\label{library-authoring-and-automatic-coverage}
\addcontentsline{toc}{subsection}{5.5 Library Authoring and Automatic Coverage}

Tiers 1--3 form the annotation-free application path. Ordinary types, domain operations and target declarations provide the inputs to elaboration. The compiler generates the obligations and, for registered Tier 3 constructions, finds and applies the reusable laws. An unsupported construction produces a coverage diagnostic. Extending that coverage would be the work of individuals versed in the domain, and would be community contribution to framework or domain-library development. We anticipate that much of that work will be Braidpoint work, and that the body of proofs that develop will serve as a roadmap that others with the interest and background to contribute.

We anticipate the general proof design case where a library author establishes a general theorem, its supported construction and its semantic binding to the PSG. Typed quotations can describe the parameters, premises and conclusion. A registered Rocq theorem offers another authoring route when its interpretation is connected to the relevant Clef operations. Both authoring routes give the compiler a reusable law whose instances it can derive from application structure.

Each admitted law should serve an actual construction and bring positive and negative applicability cases. Existing mechanized foundations reduce the work of developing each extension. As coverage grows, Tier 3 applications and recognized Tier 4 constructions can reuse that investment through ordinary program structure. Our Clef Proofs interface makes the generated instance, supporting evidence and remaining obligations inspectable beside the source; displaying the proof is independent of performing the check.

\section*{6. Integrity Across Proof and Execution Boundaries}\label{integrity-across-proof-and-execution-boundaries}
\addcontentsline{toc}{section}{6. Integrity Across Proof and Execution Boundaries}

The four tiers become useful together when evidence survives a change in how a value is represented, owned or computed. We first state the boundary correspondence, then the proof-composition rule, and finally apply both to a distributed reduction. This connects the developer\textquotesingle s freedom to partition and place work to the exact conditions the compiler and libraries must establish.

\subsection*{6.1 BAREWire and Integrity at Both Endpoints}\label{barewire-and-integrity-at-both-endpoints}
\addcontentsline{toc}{subsection}{6.1 BAREWire and Integrity at Both Endpoints}

A boundary has two native interpretations. The sender may hold a value in a register, a local region or shared memory; the receiver may use a different width, layout or execution substrate. BAREWire\textquotesingle s contract identifies how the represented value crosses that boundary and how the receiving computation interprets it. \emph{Unboxed} means that the value can have its declared concrete representation without a language-runtime object wrapper. The proof account resides in the retained semantics and admitted operations. These facts justify the endpoint operations that encode, accept and consume the unboxed value.

Let \(c\) identify an admitted boundary contract. Write \(\operatorname{Rep}_s^c(v,m_s)\) and \(\operatorname{Rep}_r^c(v,m_r)\) for the sender and receiver states representing the same abstract value \(v\). For a value-preserving codec pair, its representation theorem has the form

\[
\begin{aligned}
&\operatorname{Rep}_s^c(v,m_s)
 \land \operatorname{encode}_s^c(m_s)=b\\
&\qquad\land\operatorname{decode}_r^c(b)=\operatorname{Some}(m_r)
 \ \Longrightarrow\ \operatorname{Rep}_r^c(v,m_r).
\end{aligned}
\]

The predicates include the relevant carrier, dimensions, domain and representation choices. The theorem lets a property established about \(v\) justify the corresponding receiving value. It permits different native layouts with a proved mapping between them. A deliberately lossy conversion instead establishes a relation between the two interpreted values, with its conversion error carried into the consumer\textquotesingle s contract (§7.3).

The complete boundary rule also establishes the resources needed to use \(m_r\). A shared-memory path relates access permissions and the lifetime of common storage. A copied or networked path relates the received payload to its decoding, allocated storage and permitted consumer. Zero-copy is available when the backing storage, layout and ownership conditions support it. Across all three cases, the job, endpoint, contract version and relevant value identities connect the received instance to the established claim.

The compiler can discharge a static correspondence once while generating the runtime acceptance operations required for future inputs. An open receiver checks the conditions that incoming bytes must satisfy. The protocol proof relates an accepted payload to the correct sending event and state. Under an adversarial transport model, an origin or tamper-resistance claim additionally uses the selected authentication theorem. The boundary\textquotesingle s value interpretation and ownership requirements remain the same whether authentication is required by the deployment or the channel model already supplies the needed integrity premise.

This is how proofs carry integrity through an unboxed interface: the sender\textquotesingle s established property becomes usable by the receiver through a checked representation and resource correspondence. A hash can bind evidence to the relevant artifact or contract; the semantic theorem establishes what the bound artifact means. The receiving computation\textquotesingle s contract follows from that theorem and its local premises.

\subsection*{6.2 Mode Shifts and Proof-Term Reuse}\label{mode-shifts-and-proof-term-reuse}
\addcontentsline{toc}{subsection}{6.2 Mode Shifts and Proof-Term Reuse}

The arithmetic and protocol parts of a program pose different questions of the same computation. A bound can establish that an accumulator is large enough; an ownership judgment establishes which may update it; a relational judgment connects that update to a specified result. Mode shifts give these proofs explicit interfaces through which an established result can become a lemma in a receiving derivation. Our \href{https://arxiv.org/abs/2606.02854}{Fixed-Point Scaffolding §5} \hyperref[ref-haynes-fps-2026]{{[}26{]}} develops the categorical account. Here we connect that account to compiler-generated proof terms and their reuse.

Hăvărneanu\textquotesingle s mode theory specifies permitted dependencies through a preorder: \(m\ge k\) allows a judgment at mode \(k\) to depend on assumptions at mode \(m\). Its structural signature specifies weakening and contraction, the permissions to discard or reuse assumptions. These permissions matter to the compiler because a reusable fact about an immutable value and exclusive ownership of live storage have different rules of use. Our mode interfaces retain the corresponding contexts when those forms of evidence participate in one proof.

In \emph{Classical Adjoint Logic}, Theorems 3.18 and 4.20 \hyperref[ref-havarneanu-adjoint-2026]{{[}10{]}} establish the adjunction between downward and upward shifts for an allowed mode comparison. For shifts \(F\) from \(m\) to \(k\) and \(G\) from \(k\) to \(m\), the entailment correspondence is

\[
F(X)\preceq_k Y\quad\Longleftrightarrow\quad X\preceq_m G(Y).
\]

Here \(\preceq\) denotes derivability: a proof using the shifted premise \(F(X)\) corresponds to one whose conclusion is shifted as \(G(Y)\). The associated unit and counit derivations satisfy the triangle equations, giving the two shifts coherent composition laws. Theorem 4.20 states this result within declared compositional interfaces containing the relevant formulas. These laws specify how proofs cross an interface while retaining the structural discipline of their contexts.

Our instantiation gives each interface a source judgment, a receiving judgment, their semantic correspondence and the rules governing its premises and resources. For an interface using an adjoint pair, its registered translations and unit/counit laws form part of the checking justification. For the distributed sum, an arithmetic certificate supplies an accumulator-capacity fact; a resource rule uses that fact together with permission to update the same live accumulator. For a direct Tier 2 to Tier 4 application, the arithmetic result instead supplies a premise of the registered relational rule, as in the sampler example in §5.4. Both applications reuse the checked arithmetic evidence through the interface appropriate to the receiving proof.

An ordinary non-resource-consuming lemma application illustrates how the receiving derivation uses its evidence. For a permitted axiom basis \(A\), a checked theorem and its instantiated premise give

\[
\frac{
 A\vdash L:\forall p.\;P(p)\rightarrow Q(p)
 \qquad A;\Gamma\vdash e:P(v)
}{
 A;\Gamma\vdash L(v,e):Q(v)
}.
\]

The term \(L(v,e)\) records the theorem, its actual parameters and the evidence for its premise. Its conclusion remains derived under the same foundation and use-site hypotheses. A change of reasoning discipline additionally uses the checked translation specified by the mode interface. Resource-sensitive applications use context splitting, framing and ownership rules to account for each resource used or transferred.

For a cvc5 certificate, an accepted reconstruction or a checker with an admitted soundness justification makes the arithmetic conclusion available as a lemma to the receiving proof system. A Rocq derivation then retains that lemma\textquotesingle s checking dependencies along with those of its other premises. Imported theorems satisfy the declared foundational axiom policy; environment hypotheses, such as a device transition or a host transaction guarantee, remain conditions on the application result. This lets the compiler reuse substantial proofs while checking each new instance against its actual context.

Proof transport also has to agree with program lowering. One coordinate records the reasoning mode; the other records the compilation stage. When a source-level bound supports a resource judgment that is then lowered to a native access, the resulting evidence must justify that actual access and its target layout. Where applying the mode shift before or after lowering describes the same construction, the corresponding translations require a checked agreement under the declared semantic or proof equivalence. Our retained participant identities and preservation rules bind both routes to the same operation. The following proposition states the composition principle for such checked routes.

\textbf{Proposition 6.1 (Conditional composition across modes and lowering).} Fix an interpretation of the permitted foundation \(A\) and execution hypotheses \(\Gamma\). Let \(J_0,\ldots,J_n\) be a finite route of judgments, each indexed by its program artifact, compilation stage and reasoning mode. Resource contexts and quantitative parameters are part of a judgment. Suppose:

\begin{enumerate}
\def\labelenumi{\arabic{enumi}.}
\tightlist
\item
  Accepted evidence establishes \(J_0\) under \(A;\Gamma\).
\item
  Every edge has an admitted transport rule \(T_i\), a checked application to the actual participants of \(J_i\), and discharged side conditions \(H_i\). The rule has a sound semantic interpretation: validity of \(J_i\) and \(H_i\) entails validity of \(J_{i+1}\). Its context rules account for any resource consumption, framing or reuse.
\item
  Adjacent applications agree on the intermediate artifact, judgment interpretation and shared premises. When alternative routes are identified, their transports satisfy the declared semantic or proof equivalence.
\item
  A terminal adequacy rule relates \(J_n\) to the claimed property \(Q\) of the identified target artifact under \(\Gamma\).
\end{enumerate}

Then the composed evidence establishes \(Q\) for that target artifact under the same foundation and execution hypotheses.

\textbf{Proof.} The initial evidence supplies validity at the first position. At each edge, its discharged side conditions and the soundness of the admitted transport establish the next judgment. Matching intermediate interpretations permits induction over the finite route, including edges that change reasoning mode. Applying terminal adequacy yields \(Q\). Resource premises are transported only through the admitted context rules. Any error or failure budget remains an explicit parameter of the judgments and is transformed by those same rules; the induction adds no uncertainty to a proved judgment. \(\square\)

The primitive, mode-interface and artifact-adequacy theorems supply the premises of this composition rule. A library may establish those instances once and expose their composition as a parameterized rule. For a correctness claim about completed executions, the realization relation must ensure that every admitted target completion corresponds to a permitted source completion with the stated observation; source executions having some matching target execution is insufficient. The following example specifies one such evidence path and its conclusion.

\subsection*{6.3 Preserving a Sum Across Distributed Workers}\label{preserving-a-sum-across-distributed-workers}
\addcontentsline{toc}{subsection}{6.3 Preserving a Sum Across Distributed Workers}

Our join permits workers to evaluate disjoint partitions concurrently and valid payloads to arrive in different orders. Its result is one final rounding of an exact represented sum. The capacity law makes every admitted partial sum safe; the ownership and protocol laws associate each accepted contribution with the correct snapshot and partition once. BAREWire then carries that contribution into the receiver\textquotesingle s interpreted accumulator. The following derivation makes the freedom to change scheduling and partitioning a consequence of those combined laws.

Consider a fixed input snapshot for job \(j\), with indices \(I=\{1,\ldots,N\}\) and finite represented terms \(t_i=\delta k_i\). Here \(\delta>0\) is a declared accumulator quantum carrying the common physical dimension \(u\), and \(k_i\in\mathbb Z\). Term formation must specify whether each term is a represented input or an exact product of represented inputs. Its interpretation includes any earlier input rounding. Fix an output format \(F\) and rounding mode \(r\). The final-conversion contract covers every reachable sum, specifies any underflow behavior, and excludes overflow and unsupported exceptional outputs by a checked domain condition. The reference computation is

\[
\operatorname{Ref}_{F,r}(j)
 =\operatorname{round}_{F,r}\!\left(\delta\sum_{i\in I}k_i\right).
\]

This is one final rounding of an exact sum. A sequential fold that rounds at each addition has a different specification. Rounded or compensated worker results enter through an error relation that bounds their difference from the exact partial sums (§7.3).

\textbf{Local evidence and one mode interface.} Choose a signed accumulator width \(w\ge2\) and the conservative magnitude limit \(K=2^{w-1}-1\). Tier 1 establishes the common dimension and typed term construction. Tier 2 supplies evidence for an integer bound \(B\in\mathbb Z_{\ge0}\), the admitted term bounds \(|k_i|\le B\), and \(NB\le K\). With \(N\) and \(w\) fixed by this instance, the capacity condition is linear integer arithmetic. For a reusable implementation, a checked term-construction rule or invariant establishes the bound for every term admitted to the sum.

A finite-sum lemma then establishes

\[
\left|\sum_{i\in U} k_i\right|\le |U|B\le NB\le K
\quad\text{for every }U\subseteq I.
\]

The lemma is proved once at library level and instantiated with the term identities and capacity evidence. It covers every partial sum of distinct terms. For a selected primitive, let \(C_w(s,t)\) be the quantifier-free condition that \(s\), \(t\) and \(s+t\) all lie in \([-K,K]\). A separately established arithmetic/resource adapter \(L_{\mathrm{merge}}\) yields the mode interface

\[
\frac{
 A;\Gamma\vdash c:C_w(s,t)
 \qquad A\vdash L_{\mathrm{merge}}:\operatorname{SoundMerge}_w
}{
 A;\Gamma\vdash
 \{\operatorname{Own}(a,\operatorname{enc}_w(s))
   *\operatorname{Own}(b,\operatorname{enc}_w(t))\}
 \quad\operatorname{merge}_w(a,b)\quad
 \{\operatorname{Own}(a,\operatorname{enc}_w(s+t))
   *\operatorname{Own}(b,\operatorname{enc}_w(t))\}
}.
\]

The specified primitive reads the second accumulator and updates the first, retaining both resources. \(\operatorname{Own}\) includes the required layout, alignment and permissions; separating conjunction requires disjoint owned storage. \(\operatorname{SoundMerge}_w\) connects the actual operation to decoded integer addition with this resource behavior. The arithmetic certificate supplies a persistent pure premise. The merge judgment combines it with ownership of both accumulators and the checked law for the actual primitive.

For this example, select kernel-checked reconstruction under \(A\) as the certificate acceptance path. A solver result can support the construction of \(c\) only through an accepted reconstruction into that foundation; an unsupported reconstruction remains unresolved. The resource adapter and subsequent protocol and realization laws must also be admitted under \(A\). This acceptance path keeps the arithmetic, resource, protocol and realization judgments under the same declared foundation.

\textbf{Protocol composition.} Let \(\mathcal P\) name a finite family of disjoint, exhaustive partitions \(I_p\) of the same snapshot. An admitted worker law establishes that its decoded payload equals \(\sum_{i\in I_p}k_i\): a local invariant tracks the subset of processed indices, and sending requires complete partition coverage with each index included once. Workers transmit those exact accumulator states through the BAREWire correspondence in §6.1. Its sender and receiver predicates both decode to the same integer partial sum and physical quantum, even when their native layouts differ. The protocol admits a result only for its job, epoch and partition identity, and at most once. Its resource and acceptance laws preserve

\[
q=\sum_{p\in\operatorname{Accepted}(j)}
       \sum_{i\in I_p}k_i.
\]

Initially the accepted set is empty and \(q=0\). Accepting a fresh valid partition performs one protocol transition: test freshness, merge the payload and record acceptance. Serialization or a proved atomicity rule preserves that transition, so concurrent handlers cannot each accept the same partition as new. Rejecting a duplicate leaves both the accepted set and sum unchanged. If the fault model permits crash and recovery, a recovery invariant relates the durable acceptance record to its accumulator update. Finite identifiers, their reuse and retained state have corresponding protocol premises. Completion is enabled only when every partition in \(\mathcal P\) has been accepted.

At completion, partition disjointness and coverage give \(q=\sum_{i\in I}k_i\). The final conversion law therefore gives \(\operatorname{Ref}_{F,r}(j)\). An Aneris/Trillium-style Tier 4 refinement can connect the distributed implementation\textquotesingle s completed observations to this protocol invariant. Compiler-relational preservation then connects those observations to the target artifact. Together these instantiate Proposition 6.1. Once the construction and its adapters are established, the library can expose both the Tier 4 refinement and its Tier 3 completed-result consequence. Either application can use supported premises obtained from the PSG and checked mode compositions, retaining the same foundational dependencies.

Message loss can prevent completion without changing the proved value of a completed accepted join. Eventual completion requires a separate theorem with the applicable scheduling, delivery and failure assumptions. For approximate arithmetic, the composed reference relation states the output error budget obtained from the worker and conversion bounds.

Fidelity.Platform supplies the device and execution premises, while BAREWire supplies layout and transfer contracts. A hosted workflow, native actor, Firecracker guest or MCU needs its corresponding realization proof. Islaris \hyperref[ref-islaris]{{[}27{]}} provides a related machine-code/MMIO verification route for its supported ISA models; another processor and peripheral set requires that target\textquotesingle s instruction and peripheral semantics, together with correspondence proofs. The same result can therefore serve several execution substrates through their respective realization laws.

As described in our {\textquotedblleft{}Backend} Lowering {Architecture\textquotedblright{}} entry in our specification, Clef and Composer carry the settled PSG through Alex and target representations to native CPU, GPU, NPU and other machine artifacts, without translating the computation through C or a vendor source language. Partitioning, memory placement, synchronization and transfer remain compiler obligations along that route. For example, GPU lowering produces a native code object, while an NPU realization assigns work and storage to tiles connected by explicit queues and transfers. The compiler generates target-specific proof obligations for instruction behavior, memory access and synchronization. Before admitting the native realization, it checks evidence that the emitted operations preserve the established result and resource contracts. Device dispatch enters through its declared interface, so the same boundary discipline applies when launching the artifact and accepting its result.

A network-attached FPGA illustrates a further composition. A host hands off a computation request via eBPF, the device performs the arithmetic, and the receiving host continues with the returned result. The ingress, arithmetic and return contracts bind the job identity, byte layout, dimensions, resources and error allowance to that request. Host packet processing and device arithmetic are distinct participants in the proof. Retransmission retains the request\textquotesingle s identity; the acceptance rule prevents a repeated response from authorizing a second state update. The target consumes unboxed values under the checked representation contract, and the result enters the continuation with its established meaning.

This construction makes integrity a source of optimization. In comparing C++ with C, Stroustrup describes {\textquotedblleft{}negative} overhead {abstraction\textquotedblright{}:} richer program structure gives the optimizer information with which to reduce execution work \hyperref[ref-stroustrup-negative-overhead-2026]{{[}28{]}}. We apply that reasoning to automated proof infrastructure. Retained bounds, support and ownership facts justify eliminating redundant checks, excluded coefficient operations and unnecessary boxing. A proof established during compilation serves the executions covered by its premises; BAREWire carries the resulting contract into the target\textquotesingle s native representation. The runtime path retains the selected arithmetic, data movement and checks required by the application and boundary contracts. The same information that establishes integrity therefore also removes work from the emitted computation.

\subsection*{6.4 Principled Marshaling for C and C++ Libraries}\label{principled-marshaling-for-c-and-c-libraries}
\addcontentsline{toc}{subsection}{6.4 Principled Marshaling for C and C++ Libraries}

C and C++ enter this architecture through explicit library interfaces. A native call sends values into another representation and returns values whose meaning the caller must preserve. A C interface, including a C-facing wrapper around a C++ library, makes the boundary concrete. Consider this schematic sample-source interface:

\begin{lstlisting}
int read_samples(sample_source *source,
                 float *destination,
                 size_t capacity,
                 size_t *written);
\end{lstlisting}

For the example, the target ABI declares \lstinline!float! as IEEE binary32 and supplies the width and signedness of \lstinline!size_t!. The library\textquotesingle s contract states that a successful call initializes a prefix of \lstinline!destination!, writes its length to \lstinline!written!, and returns a success status. The permitted errors, effects on \lstinline!source! and resource behavior are also parts of that contract. A C++ wrapper can expose this interface with C language linkage while keeping the C++ object\textquotesingle s representation behind the opaque \lstinline!sample_source! handle. Its contract specifies object creation, use and release, and how any permitted C++ exception is translated to the declared error path.

Suppose the Clef caller owns storage for \(C\) elements and passes \(n\) as this call\textquotesingle s \lstinline!capacity! argument. Tier 2 generates coverage conditions for that actual count and destination, including

\[
0\le n\le C,
\qquad n\,s_{\mathrm{elem}}\le E_{\mathrm{accessible}},
\]

where \(s_{\mathrm{elem}}\) is the declared element size and \(E_{\mathrm{accessible}}\) is the byte extent of that same live destination region. The realization also preserves the multiplication and ABI conversion without word overflow. With a fixed element size, the mathematical extent condition is linear. Pointer formation, alignment and representation rules connect that arithmetic to the actual buffer passed to the call.

The return path has an equally specific proof obligation. On the successful branch, let \(r\) be the value written by this call through its \lstinline!written! argument. The routine\textquotesingle s contract gives \(0\le r\le n\) and initialized storage for that prefix. The receiving computation may then consume those \(r\) elements while its ownership and lifetime remain valid. A different counter with the same inferred range would not establish that it counts this call\textquotesingle s initialized values. The graph must retain the call, status branch, out-parameter, destination and consuming operation as related participants.

Dimensional meaning is carried by the binding\textquotesingle s domain contract. A pressure-sample library can return values interpreted in pascals even though C\textquotesingle s \lstinline!float! has no units. Its foreign binary32 representation is related to the caller\textquotesingle s selected numeric representation under an exact or bounded-error conversion law. Compatible layouts permit a direct view under the required access and lifetime rules. Other layouts use an admitted buffer and conversion path. Both preserve the declared relation to the abstract sample sequence, including any conversion error, without requiring a boxed object to carry its meaning.

Our {\textquotedblleft{}FFI} Boundary {Semantics\textquotedblright{}} specification states the contracts that our compiler tooling preserves along this chain. Farscape retains foreign declaration and ABI provenance. Clef\textquotesingle s Baker associates coverage and conversions with the actual arguments and receiving result. The PSG retains their participant relationships, and Alex witnesses those particular operands and published adaptations at the emitted call. Composer checks the correspondence of the obligation inventory, proof evidence and witnessed graph. This keeps the established condition attached to the operation that executes, including the conversion used by its consumer.

The application developer writes the call and consumes the typed result. Tier 2 obligations arise from its argument, representation and extent requirements. A registered Tier 3 boundary law supplies the ownership and success-path consequences from the library contract. A Tier 4 relation can state the behavior of a wrapper or alternative implementation when that relation needs project-specific formulation. The same progression applies when the buffer is later sent through BAREWire: the initialized sequence, its quantity meaning and the permission to consume it become premises of the receiving endpoint\textquotesingle s contract.

\subsection*{6.5 Dispatch, Reuse and Revalidation}\label{dispatch-reuse-and-revalidation}
\addcontentsline{toc}{subsection}{6.5 Dispatch, Reuse and Revalidation}

Our dispatcher selects a procedure from the actual theories and operations in an obligation, its PSG participants and semantic encoding, the registered law, the accepted producer/checker path, and the work budget. The rule supplies the proof structure; established PSG facts fill its premises. A supported Tier 2 result can feed a Tier 4 derivation directly through an admitted mode interface. Solver search and certificate acceptance are separate steps. An unsupported proof rule, incomplete certificate, cancellation or timeout leaves the obligation unresolved.

Proof checking validates a supplied derivation under the admitted rules. Proof discovery follows the selected analyses, registered lemmas and bounded search strategies. Proposition 2.1 establishes principal dimensional inference; Tier 4 supports additional judgments through explicitly checked rule applications. Each fragment states the meaning of those judgments and the bounds on elaboration and checking.

Evidence can be reused while its statement, participant binding and relevant dependencies remain unchanged. Lowering to a new representation requires checked transport, with fresh obligations for new width, signedness, layout or state assumptions. An integer range lemma, for example, can justify a selected word width when the representation relation and overflow conditions hold. The source lemma is reused; the connection to the realized operation must be established. Changes without an applicable transport invalidate the affected instances.

The engineering division is between generating the subject of a proof, establishing the proof and preserving its binding. Clef elaboration supplies the source participants and structural facts. Our graph passes retain their relationships while representation and layout are selected. Composer dispatches the supported obligations and theorem instances. Lowering must preserve or renew the evidence for the actual emitted operations, and BAREWire carries the applicable representation contract to the receiving side.

Dependency tracking makes that evidence reusable. Even an unchanged result may need a different proof instance when an argument, source guard, layout or lifetime changes. Conversely, a proved participant correspondence can transport an existing result without rerunning its original solver query. Prepared lemma libraries and selective revalidation make this an engineering discipline for maintaining integrity as the program changes.

\section*{7. Admissibility, Probability and Numerical Accuracy}\label{admissibility-probability-and-numerical-accuracy}
\addcontentsline{toc}{section}{7. Admissibility, Probability and Numerical Accuracy}

Our graph can retain a range enclosure or a probability distribution alongside a value\textquotesingle s structural type. The enclosure restricts the admissible values; the distribution assigns probability within that space. Structural inference determines admissibility, while probabilistic inference updates relative belief within the admissible space.

\subsection*{7.1 Principal Inference and Bayesian Selection}\label{principal-inference-and-bayesian-selection}
\addcontentsline{toc}{subsection}{7.1 Principal Inference and Bayesian Selection}

A principal dimensional solution represents all admissible specializations. For an identity function, it retains a parameter \(u\) in \(\mathrm{float}\langle u\rangle\to\mathrm{float}\langle u\rangle\). Specializing \(u\) to metres removes a free variable while discarding other valid uses. A prior favoring fewer free variables would prefer that specialization. That is a selection among valid uses; the principal scheme retains the family from which those choices are made. Equal positive weights over infinitely many ground assignments have an infinite total. A normalized prior requires a finite, positive total weight.

The coherent relationship is to retain the principal solution and apply a declared prior and likelihood to admissible choices when selection is needed. The following finite case makes the distinction explicit without restricting Clef\textquotesingle s type language to a finite set.

\textbf{Proposition 7.1 (Admissibility-preserving Bayesian selection).} Let \(\mathcal{H}\) be a finite candidate set, \(C(h)\) an exact admissibility predicate, \(w(h)\ge0\) prior weights, and \(\ell(E\mid h)\ge0\) likelihood values for evidence \(E\). If

\[Z_E=\sum_{h\in\mathcal{H}}\mathbf{1}_{C(h)}w(h)\ell(E\mid h)>0,\]

then

\[p(h\mid E,C)=\frac{\mathbf{1}_{C(h)}w(h)\ell(E\mid h)}{Z_E}\]

assigns zero probability to every inadmissible candidate. A maximum a posteriori choice with positive mass satisfies \(C\).

The claim follows directly from the indicator in the numerator. With decidable admissibility and effectively supplied exact rational prior weights and likelihood values, this finite construction is computable by enumeration. Its cost includes candidate evaluation and rational arithmetic at the required bit lengths. A symbolic principal solution may describe many candidates compactly, but efficient ranking over them requires a further algorithm.

An empty admissible subset means that no offered candidate in \(\mathcal H\) satisfies the structural requirements. Establishing that the source constraints themselves are inconsistent requires a complete candidate domain or a separate unsatisfiability result. A nonempty admissible subset with \(Z_E=0\) indicates a conflict or gap in the prior, likelihood or evidence. The compiler reports an empty admissible set and missing probability mass as distinct conditions. Resolving either condition preserves the dimensional and range requirements already established for the program.

Continuous distributions require a declared measure and normalization conditions. A point mass expresses a deterministic choice. An interval refinement, a probability distribution, and an unresolved inference variable carry different information, although they may constrain the same value. Our blog post \href{https://clef-lang.com/blog/deferred-inference/}{{\textquotedblleft{}The} Gift of Deferred {Inference\textquotedblright{}}} applies this distinction to compiler commitment: retain the choices still permitted by the evidence, then select a representation when the required context is available.

\subsection*{7.2 Shared Structure and Description Length}\label{shared-structure-and-description-length}
\addcontentsline{toc}{subsection}{7.2 Shared Structure and Description Length}

Conditional Kolmogorov complexity describes the shortest program for an object when specified background information is available. This provides a useful account of reusable structure in Clef. A domain library can supply laws and their accepted proofs, while a use site records a theorem identifier, parameters, and evidence for the instantiated premises. A polymorphic dimensional scheme likewise states a relationship shared by many valid uses. The algorithmic-information account \hyperref[ref-grunwald-vitanyi-2010]{{[}29{]}} makes the role of this shared context explicit.

For a fixed computable decoding convention, let \(L\) encode the available library and let a self-delimiting application description \(a\) reconstruct a particular evidence object \(e\) from \(L\). Then

\[K(e\mid L)\le |a|+c,\]

where \(c\) depends on the decoder and reference machine. The application supplies a concrete upper bound on description length. Comparing different libraries also requires accounting for their descriptions. Reconstructing and checking the evidence has a computational cost of its own.

The compilation account in \href{https://arxiv.org/abs/2606.02854}{Fixed-Point Scaffolding} gives this reuse an additional requirement: the instantiated law must continue to justify the corresponding operation as its representation or reasoning mode changes. Shared structure can reduce what an application must specify while preserving the premises that make the result valid. These concrete descriptions make reuse measurable: under the declared encoding, the application description and fixed decoder constant supply the displayed upper bound, while reconstruction and checking contribute to compilation cost.

Solomonoff\textquotesingle s universal semimeasure \hyperref[ref-solomonoff-1964]{{[}30{]}} assigns weight to programs that produce an observed prefix:

\[M(x)=\sum_{p:\,x\preceq U(p)}2^{-|p|},\]

with the usual coding conditions on the reference machine. It is lower semicomputable: progressively running candidate programs supplies increasing lower bounds. It is not generally computable. Leike and Hutter \hyperref[ref-leike-hutter-2015]{{[}31{]}} distinguish this object from its normalized and conditional variants.

Dimensional admissibility restricts which candidates may be selected. An effective prior specifies how those candidates are represented and how their weights are computed. For an infinite candidate class, normalized probabilities require an effective procedure for the normalizing sum, or certified bounds sufficient for the declared approximation.

Description-length and Bayesian objectives can correspond when a particular code, model class, and likelihood are specified. A prefix-code length counts the bits in a self-delimiting candidate description under the declared encoding. For dimensional assignments, that description specifies the chosen measures and exponents relative to the shared declarations. A description-length selection rule therefore defines this encoding and its relationship to the probabilistic objective.

\subsection*{7.3 Carrying Accuracy Bounds Across Distributed Workers}\label{carrying-accuracy-bounds-across-distributed-workers}
\addcontentsline{toc}{subsection}{7.3 Carrying Accuracy Bounds Across Distributed Workers}

The exact join in §6 uses a quire or another covered exact accumulator to separate partitioning from rounding. Its fixed width determines storage; the term bounds and partial-sum law establish capacity. One final conversion then supplies the specified output rounding. When workers perform approximate calculations, the same boundary and protocol discipline can instead preserve an explicit accuracy and failure contract.

Urschel\textquotesingle s elimination-growth results supply reusable bounds for linear-system solvers \hyperref[ref-urschel-growth-2026]{{[}32{]}}. His Theorem 1.2 gives, for any fixed nonsingular real \(n\times n\) matrix \(A\), Haar-distributed orthogonal \(Q\), and \(0<\delta<1\), a bound for elimination without pivoting:

\[
\Pr\!\left[\operatorname{growth}(QA)>
 \frac{14n^{3/2}\sqrt{\log(en)}}{\delta}\right]<\delta.
\]

This is a new quantitative bound for an established preconditioning approach. His Gaussian-input partial-pivoting result, Theorem 1.4, has a different matrix law. The Haar route is useful here because the matrix may come from a structured domain computation: its nonsingularity and the fresh random preconditioner supply the relevant premises. These premises can be stated for a linear-system solve inside a simulator, controller or distributed service.

A growth bound informs the range and precision needed for a solve when combined with the transformed-input scale, arithmetic-error analysis and output sensitivity. A registered law must relate the sampler, formation of the transform and realized elimination to the stated real-arithmetic calculation and sampling law. The resulting instance can supply a worker error bound and a failure event. A representation that must remain safe on every execution additionally uses a deterministic coverage argument or a checked guard and fallback for exceptional growth. This connects probabilistic growth bounds to a usable computation contract while preserving the stronger coverage required for storage and machine operations.

The next proposition shows how such a worker law contributes to integrity across the distributed boundary. It applies equally when another theorem supplies the local bound.

\textbf{Proposition 7.2 (Accuracy and failure bounds for a completed distributed join).} Fix a finite set of worker contributions for one job snapshot. Let \(s_i\) be worker \(i\)\textquotesingle s ideal contribution, with common physical dimension \(u\), and \(z_i\) its computed value before transfer. In the selected execution and sampling model, suppose an event \(B_i\) covers violations of the worker\textquotesingle s accuracy contract, with

\[
\Pr(B_i)\le\delta_i,
\qquad B_i^c\Longrightarrow |z_i-s_i|\le\epsilon_i.
\]

The bounds apply under each scheduler admitted by that model. Suppose every accepted BAREWire payload has receiving interpretation \(y_i\) satisfying \(|y_i-z_i|\le\eta_i\). Use the identity, coverage, ownership and atomic acceptance laws of §6, so a completed join includes each required contribution once. Let the receiver accumulate the represented \(y_i\) exactly within its established capacity and let final conversion introduce error at most \(\eta_{\mathrm{out}}\). For fixed nonnegative bounds, put

\[
S=\sum_i s_i,
\qquad
E=\sum_i(\epsilon_i+\eta_i)+\eta_{\mathrm{out}}.
\]

Then its completed output \(\widehat S\) satisfies

\[
\Pr\!\left[\mathrm{completed}\ \land\ |\widehat S-S|>E\right]
 \le\min\!\left(1,\sum_i\delta_i\right).
\]

\textbf{Proof.} Outside \(\bigcup_i B_i\), the worker and boundary contracts give \(|y_i-s_i|\le\epsilon_i+\eta_i\). The protocol invariant supplies each contribution once, and exact receiving accumulation followed by the stated conversion gives \(|\widehat S-S|\le E\) by the triangle inequality. Thus an incorrect completed result lies in the union of the bad events. The union bound establishes the conclusion; it needs no independence between those events. \(\square\)

The value-error bounds have dimension \(u\); the \(\delta_i\) are dimensionless probabilities. An exact BAREWire mapping contributes \(\eta_i=0\). Tier 2 can check a selected allocation against the application\textquotesingle s error and failure budgets, while Tier 3 supplies the instantiated worker and boundary laws. The relational conclusion connects the realized distributed output to its reference value. Urschel\textquotesingle s theorem contributes an elimination-growth premise to that composition; ownership, message acceptance and realization preservation establish the rest of the result.

The probability in the proposition concerns an incorrect completion over the declared experiment. Conditioning on completion can change the bound if completion depends on the samples. Retransmitting the same established payload does not resample the worker computation; a fresh randomized recomputation enters the risk account as another attempt. These distinctions keep a retry or completion policy from silently changing the experiment described by the evidence.

For a threshold decision at \(\tau\), a checked margin \(|S-\tau|>E\) gives the same decision from \(S\) and \(\widehat S\) outside the combined bad event whenever the join completes. This is a useful observation relation for a distributed controller or a service returning a computed result. A relation over raw output distributions or over noncompletion behavior states its own observation and adequacy conditions. The engineering consequence is that the consumer can use the accuracy guarantee that actually survived the computation and its boundary crossings.

\subsection*{7.4 Optimizer Configuration as Design-Time Verification}\label{optimizer-configuration-as-design-time-verification}
\addcontentsline{toc}{subsection}{7.4 Optimizer Configuration as Design-Time Verification}

Parallel tangent calculations provide another use for the same discipline. Their coordinate system determines the meaning of a projection, their number determines work and storage, and their sampling law determines any statistical guarantee. These choices can become explicit premises of the optimizer contract.

Let \(\theta\) contain \(n\) potentially heterogeneous physical parameters. Choose fixed, nonzero reference scales \(s_i\) with the same units as \(\theta_i\), and set \(S=\operatorname{diag}(s_i)\), \(z=S^{-1}\theta\) and \(\widetilde f(z)=f(Sz)\). The coordinates \(z\) are dimensionless. This choice supplies a Euclidean geometry in which tangent directions can be compared; changing the reference scales can change that geometry and the optimizer (Surace et al. \hyperref[ref-surace-metric-2020]{{[}33{]}}).

For a matrix \(V\in\mathbb{R}^{n\times k}\) of linearly independent, dimensionless tangent directions in these coordinates, the projected gradient is

\[
P_V(\nabla_z\widetilde f)
 =V(V^\top V)^{-1}V^\top\nabla_z\widetilde f,
\qquad
\nabla_z\widetilde f=S^\top
 \begin{pmatrix}
 \partial f/\partial\theta_1\\ \vdots\\ \partial f/\partial\theta_n
 \end{pmatrix}_{\theta=Sz}.
\]

The right-hand column represents the differential of \(f\) in physical coordinates. The chain rule transports it to the chosen dimensionless coordinates, where Euclidean identification gives the displayed gradient. Its components all have the loss\textquotesingle s units. An update is

\[
z^+=z-\eta P_V(\nabla_z\widetilde f),
\qquad \theta^+=Sz^+.
\]

If the loss carries units, \(\eta\) must carry their inverse for this update to preserve dimensionless coordinates. A dimensionless loss admits a dimensionless \(\eta\). The reference scales, coordinate interpretation and update conversion belong to the optimizer contract.

The full-rank premise is required for the displayed inverse. Rank-deficient directions require a different specified solve. Flügel et al.\textquotesingle s multi-tangent method \hyperref[ref-flugel-multitangent-2026]{{[}34{]}} studies this construction and its estimation properties under stated assumptions.

A compiler can check the units and shapes of the products and account for the storage of directions, directional derivatives and the Gram matrix. Here \(n\) is the dimension of the differentiated parameter or activation coordinates, established from the program. Physical units do not determine it. Forward geometric sparsity can reduce work, but a reduction in gradient dimension requires a dependency or rank argument.

The ratio \(k/n\) is useful descriptive information. A probabilistic approximation or variance guarantee also depends on the tangent distribution, coordinate scaling and the theorem\textquotesingle s other premises. The unmeasured gradient component needs its own bound, as shown in \href{https://arxiv.org/abs/2603.18104}{ADM §2.4}. Exact quire accumulation can support covered Gram-matrix entries; coordinate conversion, inversion and the final update retain separate range, invertibility and error obligations.

Our Lattice readout design exposes this evidence: units, shapes, resource requirements, coordinate choices and the premises supporting any statistical bound. The readout distinguishes established conditions from choices that still need domain evidence.

ADM develops a proposed metric-aware extension that replaces the Euclidean pairing in the restricted tangent calculation by a declared probabilistic metric \hyperref[ref-haynes-adm-2026]{{[}9{]}}. Our verification interface must bind each Gram entry and directional loss observation to the same tangent occurrences, parameter state, observation map and metric. Positive semidefiniteness supplies nonnegative quadratic values, while a unique coefficient solve requires positive definiteness on the selected directions. For a singular Gram matrix, the directional loss vector must annihilate its kernel for the corresponding linear system to be consistent. A selected representative, quotient or regularized solve then needs its own contract. Regularization changes the update and must remain visible in the optimizer specification.

Estimated entries introduce further obligations: sampling assumptions and error bounds, numerical symmetry and definiteness, conditioning, solve error and the resulting update\textquotesingle s contract. Passing a shape check or computing a finite matrix does not establish these claims. A changed output scale, tangent identity or sampler invalidates the dependent instance even when the matrix dimensions and machine representation remain identical. The full accounting includes retained samples or sufficient statistics and the cost of restoring their validity after a model update.

\section*{8. Runtime Mechanisms and Compiler Contracts}\label{runtime-mechanisms-and-compiler-contracts}
\addcontentsline{toc}{section}{8. Runtime Mechanisms and Compiler Contracts}

A program can enforce a feasible set at runtime, minimize a residual, retrieve an existing value or specialize a geometric computation. Each mechanism establishes a different property and retains different execution costs. Our verification account identifies which parts can be proved before execution and which remain obligations of the running implementation. Learning systems provide several examples within this comparison.

\subsection*{8.1 Runtime Feasibility Enforcement}\label{runtime-feasibility-enforcement}
\addcontentsline{toc}{subsection}{8.1 Runtime Feasibility Enforcement}

Projection onto a nonempty closed convex set establishes membership in that set \hyperref[ref-bauschke-combettes-2011]{{[}35{]}}. A controller or solver may use such a projection before accepting a newly computed vector. Some projections have closed forms, while others require optimization; their cost depends on the set and solver. A finite-precision implementation must establish the required membership or declared tolerance under its own arithmetic contract.

Dimensional checking can verify that the variables and constraints in a projection problem are well formed. A static proof that every output already belongs to the feasible set can justify eliminating the projection. A projection retained at runtime must satisfy its membership or tolerance contract.

\subsection*{8.2 Residual and Penalty Objectives}\label{residual-and-penalty-objectives}
\addcontentsline{toc}{subsection}{8.2 Residual and Penalty Objectives}

A residual expresses a discrepancy against a specified equation; a penalty makes that discrepancy contribute to an optimization objective. Dimensional analysis can reject incompatible quantities in the expression. Minimizing the objective encourages a small residual, while a verified accuracy or feasibility claim requires the corresponding bound or acceptance criterion.

Physics-informed neural networks use domain equations through training residuals \hyperref[ref-raissi-pinns-2019]{{[}36{]}}. Our compiler can check the residual\textquotesingle s structural meaning and preserve its representation contract through the training computation. Evaluation or a separate theorem establishes how closely the resulting approximation satisfies the domain equation. Residual evaluation remains an execution cost while it is part of the chosen objective.

\subsection*{8.3 Lookup and Selection}\label{lookup-and-selection}
\addcontentsline{toc}{subsection}{8.3 Lookup and Selection}

A lookup returns an existing entry according to a key or selection policy. A cache, table of precomputed values or remote store must also establish that the returned entry has the version, interpretation and validity required by its consumer. Our BAREWire boundary binds those conditions to the request and response, including the conditions that require runtime validation.

An external example of lookup as a source of model capacity is Cheng et al.\textquotesingle s Engram \hyperref[ref-cheng-engram-2026]{{[}37{]}}, which introduces conditional memory through scalable lookup and evaluates it against matched model budgets. For our verification account, the relevant question is how a selected entry becomes valid input to the operation that consumes it. Our consultation policy in §4.3 makes that obligation explicit for a response that updates recurrent state. The selection mechanism and the update rule each need a contract, and §10 includes lookup, conversion, validation and retained state in the cost account.

\subsection*{8.4 Geometric Specialization}\label{geometric-specialization}
\addcontentsline{toc}{subsection}{8.4 Geometric Specialization}

Geometric structure can identify arithmetic that a solver can omit while preserving its result. To make that optimization dependable, the compiler must retain what the coordinates denote and the algebraic rules under which the omitted terms vanish.

For example, a solver may represent spatial lines with Plücker coordinates, which encode direction and position. Its declared embedding specifies how those coordinates participate in geometric operations. Our compiler account requires conversion rules to preserve that interpretation and uses blade-support evidence to exclude coefficient products proved zero. The resulting specialization reduces the arithmetic performed by the solver.

A PHG-guided comparison can count the coefficient products removed, measure their share of execution time, and retain the solver\textquotesingle s embeddings and search procedure as explicit experimental conditions.

\subsection*{8.5 Cost Structure Summary}\label{cost-structure-summary}
\addcontentsline{toc}{subsection}{8.5 Cost Structure Summary}

{\def\LTcaptype{none} 
\begin{longtable}[]{@{}
  >{\raggedright\arraybackslash}p{(\linewidth - 4\tabcolsep) * \real{0.3333}}
  >{\raggedright\arraybackslash}p{(\linewidth - 4\tabcolsep) * \real{0.3333}}
  >{\raggedright\arraybackslash}p{(\linewidth - 4\tabcolsep) * \real{0.3333}}@{}}
\toprule\noalign{}
\begin{minipage}[b]{\linewidth}\raggedright
Mechanism
\end{minipage} & \begin{minipage}[b]{\linewidth}\raggedright
Establishes or supports
\end{minipage} & \begin{minipage}[b]{\linewidth}\raggedright
Costs that remain
\end{minipage} \\
\midrule\noalign{}
\endhead
\bottomrule\noalign{}
\endlastfoot
Convex projection & Membership in a specified feasible set, under the implementation\textquotesingle s arithmetic contract & Projection evaluation, unless statically proved redundant \\
Residual or penalty objective & A specified discrepancy and its contribution to an optimization criterion & Residual evaluation and evidence for the required accuracy or feasibility \\
Lookup and selection & Retrieval under a specified key or selection policy & Lookup, version handling, conversion and required validation \\
Typed elaboration & The admitted structural and proof obligations & Constraint solving, preservation checks and necessary runtime boundary validation \\
\end{longtable}
}

Eliminating a repeated check can reduce execution cost when the compiler has a valid proof covering every use. The proof must survive changes to code, context and representation. Re-elaboration after invalidation is part of that cost. Section 10 includes this revalidation work in the complete cost of maintaining the result contract.

\section*{9. Related Work}\label{related-work}
\addcontentsline{toc}{section}{9. Related Work}

The preceding constructions connect several traditions of verification. The relevant comparisons concern what each establishes, how its evidence is accepted, and which part of that evidence our compiler can preserve and reuse.

Edinburgh LCF separated programmable proof search from trusted theorem construction. Its abstract theorem type protected construction through the logic\textquotesingle s admitted inference operations; tactics could organize proofs while that boundary governed acceptance (Gordon \hyperref[ref-gordon-lcf-hol-2000]{{[}38{]}}). Our design uses a related separation between generating an obligation, finding evidence and admitting that evidence under the selected interpretation. The additional task here is to preserve its binding across several compiler judgments and execution boundaries. Principal type inference is a distinct contribution of the ML language tradition, represented by Hindley and Milner\textquotesingle s work.

Kennedy\textquotesingle s units-of-measure work establishes principal inference for dimensional equations in an ML-family type system. Clef retains those facts into the phases that select representations and layouts. Petricek, Orchard, and Mycroft\textquotesingle s coeffects \hyperref[ref-petricek-coeffects-2014]{{[}39{]}} provide a vocabulary for requirements a computation places on its context.

Fu, Angiuli and Tobin-Hochstadt \hyperref[ref-fu-occurrence-2026]{{[}5{]}} examine occurrence typing, where restricted term references inside types interact with substitution and scoping. They identify soundness defects and give a semantic repair formalized in Lean. Their result concerns that calculus; its relevance here is the need to preserve a fact\textquotesingle s meaning when bindings or available information change. Clef keeps its ordinary dimensional inference within the selected measure algebra and assigns richer value dependencies to graph analyses and checked interfaces. Each graph transformation must therefore preserve both the retained fact and its binding to the relevant operation.

Clef and Composer also draw on OCaml\textquotesingle s module discipline, Scheme\textquotesingle s nanopass methodology, and MLKit\textquotesingle s region analysis. F* and Dafny inform the integration of automated verification into a programming workflow. F\# contributes practical experience with Python-influenced syntax, mailbox processors, quotations and active patterns.

Our compiler derives proof obligations from the computation\textquotesingle s data, control, resource and communication relationships. That account applies to supported ordinary functions and parallel tasks as well as explicit actor programs. Actor semantics makes local state and message boundaries particularly clear, providing a useful illustration of how the same structure extends across a network. Erlang\textquotesingle s process isolation and asynchronous message passing give this approach an established engineering lineage; Armstrong dates the start of Erlang\textquotesingle s development to 1986 \hyperref[ref-armstrong-erlang-2003]{{[}40{]}}.

We anticipate that mechanized proof libraries will supply reusable reasoning about those relationships. Iris \hyperref[ref-jung-iris-2018]{{[}16{]}} provides ownership and invariant rules for concurrent state. Aneris \hyperref[ref-aneris]{{[}18{]}} brings that resource discipline to distributed execution, while Trillium \hyperref[ref-timany-trillium-2024]{{[}22{]}} relates implementation traces to abstract protocols. The Verdi precedent \hyperref[ref-wilcox-verdi-2015]{{[}41{]}} contributes verified system transformations under explicit fault assumptions. Islaris \hyperref[ref-islaris]{{[}27{]}} brings Iris-based reasoning to machine code for its supported instruction-set models, including memory-mapped I/O (MMIO) interactions governed by an explicit device protocol. Applying that route to Clef retains the target semantics and correspondence obligations of §6.3. Our library authors bind the applicable rules and theorems to Clef operations, BAREWire contracts and target realizations. The compiler then instantiates those laws from the retained graph and discharges their supported local premises. RefinedC \hyperref[ref-refinedc]{{[}42{]}} provides a precedent for automating foundational verification; our application workflow derives its obligations from the PSG. This makes established proof work available to ordinary programs through reusable compiler rules.

McErlang provides a verification precedent within the Erlang lineage. Its controlled runtime explores communication, scheduling and fault behavior; safety monitors and temporal-logic specifications express properties of those executions. In the supervisor case study, state-space growth required selecting tractable scenarios instead of verifying the component in full (Castro et al. \hyperref[ref-castro-mcerlang-2011]{{[}43{]}}).

We carry that verification goal further through the compositional proof infrastructure of Iris \hyperref[ref-jung-iris-2018]{{[}16{]}} and Aneris \hyperref[ref-aneris]{{[}18{]}}. Iris supplies reusable rules for ownership and invariants; Aneris uses those foundations to reason about communicating participants and the distributed system they form. The advance for our compiler account is coverage by a parameterized theorem: its proof can establish a protocol property for every configuration and execution satisfying its premises. Once that law and its semantic binding are registered, the compiler obtains the actual participants and parameters from the PSG and checks the remaining application conditions. This gives supported programs the benefit of the general distributed proof through ordinary compilation. Checked mode interfaces connect the theorem to local arithmetic and resource evidence, while preservation rules carry the result through BAREWire boundaries and native lowering. The engineer can therefore vary scheduling, partitioning and message order within the theorem\textquotesingle s assumptions while retaining the established guarantee.

Learning systems provide a further application. Categorical deep learning studies algebraic composition of learning architectures (Gavranovic et al. \hyperref[ref-gavranovic-categorical-2024]{{[}44{]}}); using such a representation in Clef requires an interpretation of the programs and their updates. Forward gradients and posit quires supply separate derivative and accumulation mechanisms, with distinct memory and arithmetic obligations. Our \href{https://arxiv.org/abs/2603.18104}{ADM paper} develops that application. It depends on the structural facts, protocol laws and realization relations developed here.

\section*{10. Engineering Direction and Evaluation}\label{engineering-direction-and-evaluation}
\addcontentsline{toc}{section}{10. Engineering Direction and Evaluation}

Our engineering direction is one compiler workflow for generating obligations, applying registered laws and retaining their justification through realization. The application-level measure of success is broader coverage without a corresponding increase in developer design-time proof authoring or annotation. Tiers 1--3 derive their evidence from ordinary program and domain structure; Tier 4 identifies project-specific relations until registered constructions provide enough structure for automatic recognition.

The principal-inference result concerns the measure fragment. Blade support, lifetimes, ranges and capabilities retain their own interpretations and transfer rules. Their composition must have a termination account for the chosen analysis. Arithmetic admission likewise identifies the actual theory: nonlinear real polynomial arithmetic has decision procedures (Tarski \hyperref[ref-tarski-1951]{{[}45{]}}), while general integer polynomial satisfiability is undecidable (Matiyasevich \hyperref[ref-matiyasevich-1970]{{[}46{]}}). A work budget controls an implementation\textquotesingle s response time independently of a theory\textquotesingle s decidability. These distinctions guide dispatch and the diagnostics shown at a commitment boundary.

\subsection*{10.1 Assessing Integrity Across the Chain}\label{assessing-integrity-across-the-chain}
\addcontentsline{toc}{subsection}{10.1 Assessing Integrity Across the Chain}

The distributed join supplies a common assessment from source to receiver. The same declared computation should support several partition maps, arrival orders and target representations under their corresponding laws. A completed result must have the stated exact or bounded-error relation to its reference. The following comparisons isolate the mechanisms that establish that result.

{\def\LTcaptype{none} 
\begin{longtable}[]{@{}
  >{\raggedright\arraybackslash}p{(\linewidth - 4\tabcolsep) * \real{0.3333}}
  >{\raggedright\arraybackslash}p{(\linewidth - 4\tabcolsep) * \real{0.3333}}
  >{\raggedright\arraybackslash}p{(\linewidth - 4\tabcolsep) * \real{0.3333}}@{}}
\toprule\noalign{}
\begin{minipage}[b]{\linewidth}\raggedright
Assessment
\end{minipage} & \begin{minipage}[b]{\linewidth}\raggedright
Required positive evidence
\end{minipage} & \begin{minipage}[b]{\linewidth}\raggedright
Discriminating change
\end{minipage} \\
\midrule\noalign{}
\endhead
\bottomrule\noalign{}
\endlastfoot
Automatic Tier 2/3 coverage & Ordinary reduction and protocol code generates the applicable conditions and lemma instance & A construction outside the registered law produces a specific coverage diagnostic \\
Argument and boundary binding & The source value, actual call operand, representation conversion and receiver use share a justified correspondence & Changing the actual operand or receiving contract invalidates the affected instance even if an arithmetic answer is unchanged \\
Concurrent acceptance & Reordering and duplicating arrivals preserves the accepted-set and accumulator invariants & Separating freshness testing from the atomic acceptance transition exposes the duplicate-admission obligation \\
Relational realization & Each admitted partition and target implementation preserves the selected observation relation & A changed rounding order, width, signedness or emitted operation requires the corresponding new relation or evidence \\
Error and failure budgets & Worker error, boundary conversion and failure events compose to the application contract & A fresh randomized retry or altered observation policy changes the relevant risk account \\
Direct composition across tiers & A checked Tier 2 bound becomes a premise in a Tier 4 proof term through its registered interface & Rebinding the bound to another value or removing a receiving-rule premise invalidates the application \\
Evidence reuse & Unchanged dependencies reuse an accepted instance; checked transports preserve established facts & Changed source support, theorem assumptions or target dependencies invalidate only their affected consumers \\
Geometric observation & Metric evidence names the same program, statistic, coordinates and sampling law as its consumer & Deleting an unused parameter-dependent random draw requires renewed trace-geometry evidence even when output laws agree \\
Degenerate metric solve & A singular restricted system has a loss vector compatible with its kernel and an admitted choice of solution & Replacing the loss vector by one nonzero on that kernel rejects an exact solve, despite unchanged dimensions \\
Information and recovery & A reconstruction claim supplies a recovery map and the required inverse equations & Dropping an independent parameter-free fair coin can lose no parameter information while making its realized outcome unrecoverable \\
\end{longtable}
}

These comparisons distinguish a local solver result from a proof about the delivered computation. An artifact check binds to the exact graph, query, operations and output under review. A relational theorem establishes the property that may be transported. Together they make an incorrect boundary mapping or stale proof visible at the point where it would otherwise justify a receiving operation.

\subsection*{10.2 Authoring Cost and Compiler Cost}\label{authoring-cost-and-compiler-cost}
\addcontentsline{toc}{subsection}{10.2 Authoring Cost and Compiler Cost}

The cost account separates general library development from repeated application use. A useful evaluation records which Tier 3 laws the compiler recognizes without project annotations, which Tier 4 relations still need explicit input, and which later become inferable from the hypergraph. Each expansion of coverage should report the construction it supports and the compiler evidence that determines applicability.

For the compiler, measure obligation generation, analysis, solving, certificate processing, theorem application and final validation. Assess mode-shift reuse by comparing derivations that reuse checked lemmas with derivations that rebuild the same premises under the same foundation and result contract. Account for the initial evidence construction, certificate reconstruction, interface and kernel checking, and dependency revalidation across repeated uses. These measurements should identify the constructions and reuse patterns for which the initial investment is amortized. The direct Tier 2 to Tier 4 case should construct the relational proof from its actual premises, with no intermediate Tier 3 obligation unless that proof requires one.

Cold and repeated edits exercise different costs. An unrelated edit should preserve reusable evidence; a changed participant, representation or protocol premise should renew the affected obligations. The editing experience depends on this complete cycle, including cancellation and unresolved-result diagnostics.

For execution, measure worker computation, representation conversion, copying or shared-memory access, message acceptance, and retained protocol state. Compare only systems that meet the same result, failure and resource contract. Exact accumulation can trade a wider local carrier for freedom in parallel reduction and fewer reruns to diagnose rounding differences; a useful measurement counts both sides of that trade. This evaluation accounts for execution efficiency and the cost of preserving integrity across the complete result path.

\section*{11. Construction and Evidence}\label{construction-and-evidence}
\addcontentsline{toc}{section}{11. Construction and Evidence}

Our construction connects arithmetic, ownership, protocol and realization guarantees through the PSG and BAREWire\textquotesingle s unboxed boundary contracts. The worked join demonstrates the resulting freedom: disjoint work can proceed concurrently, valid partial results can arrive in different orders, and their completed combination preserves the specified sum and final rounding. The probabilistic extension carries worker and boundary error into the consumer\textquotesingle s stated accuracy and failure contract.

The four tiers organize the proof work needed to establish these properties. Structural inference and arithmetic checks supply local evidence; reusable system theorems extend it across participants; relational proofs connect computations and realizations. Mode shifts allow an established result to become a lemma wherever its conclusion is required, including directly from Tier 2 to Tier 4. Application developers obtain the coverage of Tiers 1--3 through ordinary code, while library authors supply the general proofs and Tier 4 annotations identify project-specific relations.

Urschel\textquotesingle s elimination-growth bounds \hyperref[ref-urschel-growth-2026]{{[}32{]}} and Hăvărneanu\textquotesingle s shift-adjunction results \hyperref[ref-havarneanu-adjoint-2026]{{[}10{]}} contribute complementary foundations for this account. The former supply quantitative premises that, with the scale, arithmetic-error, sensitivity and coverage arguments of §7.3, inform range and precision choices for linear-system computations. The latter supply formal laws for composing evidence across reasoning modes. Our compiler interpretation connects both to the actual operations and resources being verified. Integrity and efficiency follow the same retained information: justified bounds inform storage and arithmetic requirements, while checked proof reuse reduces repeated verification work.

Our first-release design includes differential compilation as a further consequence of this structure. A design-time edit identifies an affected scope through the PSG\textquotesingle s semantic dependencies, including joint constraint participants, proof premises and boundary contracts. Composer re-establishes the required guarantees for that scope and re-lowers the resulting fragments; unaffected fragments retain their valid evidence and target realizations. Checked contracts at the replacement boundaries justify their recomposition. This extends the negative-cost-of-abstraction argument to the development workflow: semantic information supports both safe replacement and selective recompilation, so the structure used to establish integrity also avoids repeating unaffected compilation work.

Our design connects decidable local checks with compositional distributed proofs within the compiler's verification and lowering pipeline, automatically discharging routine proof obligations while keeping their premises and evidence available for inspection. The same evidence supports precision and resource choices, proof reuse and differential compilation: each edit requires renewed verification and lowering only within its affected semantic scope, preserving valid work elsewhere to shorten the development build cycle. With Clef and Composer maintaining these contracts, we expect to provide developers with a foundation for building everyday applications and advancing accelerated and distributed systems across a variety of processor types in a way that\textquotesingle s truly fearless.

\section*{Software Availability}\label{software-availability}
\addcontentsline{toc}{section}{Software Availability}

The Clef language, Composer compiler, and supporting libraries described in this paper are developed under the Fidelity Framework project. Source repositories are available at \url{https://github.com/FidelityFramework}. The language specification, design rationale, and compiler documentation are published at \url{https://clef-lang.com}. Central components of the framework are dual-licensed; terms are detailed in each repository. The compiler design connects these components through the obligation and preservation interfaces described in this paper.

\section*{References}\label{references}
\addcontentsline{toc}{section}{References}

\leavevmode\protect\phantomsection\label{ref-kennedy-units-2009}{}{[}1{]} Kennedy, A. (2009). Types for units-of-measure: Theory and practice. In \emph{Central European Functional Programming School}, LNCS 6299. Springer.

\leavevmode\protect\phantomsection\label{ref-hindley-1969}{}{[}2{]} Hindley, J. R. (1969). The principal type-scheme of an object in combinatory logic. \emph{Transactions of the American Mathematical Society}, 146:29--60.

\leavevmode\protect\phantomsection\label{ref-milner-1978}{}{[}3{]} Milner, R. (1978). A theory of type polymorphism in programming. \emph{Journal of Computer and System Sciences}, 17(3):348--375.

\leavevmode\protect\phantomsection\label{ref-haynes-dts-dmm-2026}{}{[}4{]} Haynes, H. (2026). Dimensional type systems and deterministic memory management: Design-time semantic preservation in native compilation. \href{https://arxiv.org/abs/2603.16437}{arXiv:2603.16437}.

\leavevmode\protect\phantomsection\label{ref-fu-occurrence-2026}{}{[}5{]} Fu, Y., Angiuli, C., and Tobin-Hochstadt, S. (2026). Revisiting soundness for occurrence typing, semantically. \href{https://arxiv.org/abs/2609.16299v1}{arXiv:2609.16299v1}.

\leavevmode\protect\phantomsection\label{ref-baydin-forward-gradients-2022}{}{[}6{]} Baydin, A. G., Pearlmutter, B. A., Syme, D., Wood, F., and Torr, P. (2022). Gradients without backpropagation. \href{https://arxiv.org/abs/2202.08587}{arXiv:2202.08587}.

\leavevmode\protect\phantomsection\label{ref-haynes-phg-2026}{}{[}7{]} Haynes, H. (2026). The program hypergraph: Multi-way relational structure for geometric algebra, spatial compute, and physics-aware compilation. \href{https://arxiv.org/abs/2603.17627}{arXiv:2603.17627}.

\leavevmode\protect\phantomsection\label{ref-haynes-nft-2026}{}{[}8{]} Haynes, H. (2026). Negative and fractional types in the Fidelity Framework: A proposed operational core for reversibility and resource pairing. \href{https://arxiv.org/abs/2606.04352}{arXiv:2606.04352}.

\leavevmode\protect\phantomsection\label{ref-haynes-adm-2026}{}{[}9{]} Haynes, H. (2026). Adaptive domain models: Bayesian evolution, warm rotation, and principled training for geometric and neuromorphic AI. \href{https://arxiv.org/abs/2603.18104}{arXiv:2603.18104}.

\leavevmode\protect\phantomsection\label{ref-havarneanu-adjoint-2026}{}{[}10{]} Hăvărneanu, A. (2026). \emph{Classical Adjoint Logic}. Pre-publication manuscript, July 12.

\leavevmode\protect\phantomsection\label{ref-gustafson-posit-2017}{}{[}11{]} Gustafson, J. L., and Yonemoto, I. T. (2017). Beating floating point at its own game: Posit arithmetic. \emph{Supercomputing Frontiers and Innovations}, 4(2).

\leavevmode\protect\phantomsection\label{ref-brian-units}{}{[}12{]} Brian team. Physical units. \emph{Brian 2 documentation}. \href{https://brian2.readthedocs.io/en/stable/user/units.html}{Documentation}.

\leavevmode\protect\phantomsection\label{ref-wang-hrm-2025}{}{[}13{]} Wang, G., Li, J., Sun, Y., Chen, X., Liu, C., Wu, Y., Lu, M., Song, S., and Abbasi Yadkori, Y. (2025). Hierarchical reasoning model. \href{https://arxiv.org/abs/2506.21734}{arXiv:2506.21734}.

\leavevmode\protect\phantomsection\label{ref-roy-lambda-rlm-2026}{}{[}14{]} Roy, A., Tutunov, R., Ji, X., et al. (2026). The Y-combinator for LLMs: Solving long-context rot with lambda-calculus. \href{https://arxiv.org/abs/2603.20105}{arXiv:2603.20105}.

\leavevmode\protect\phantomsection\label{ref-wadler-1989}{}{[}15{]} Wadler, P. (1989). Theorems for free! In \emph{Proceedings of the Fourth International Conference on Functional Programming Languages and Computer Architecture}, 347--359. \href{https://doi.org/10.1145/99370.99404}{Paper}.

\leavevmode\protect\phantomsection\label{ref-jung-iris-2018}{}{[}16{]} Jung, R., Krebbers, R., Jourdan, J.-H., Bizjak, A., Birkedal, L., and Dreyer, D. (2018). Iris from the ground up: A modular foundation for higher-order concurrent separation logic. \emph{Journal of Functional Programming}, 28, e20. \href{https://www.mpi-sws.org/~dreyer/papers/iris-ground-up/paper.pdf}{Paper}.

\leavevmode\protect\phantomsection\label{ref-hinrichsen-actris-2022}{}{[}17{]} Hinrichsen, J. K., Krebbers, R., and Bengtson, J. (2022). Actris 2.0: Asynchronous session-type based reasoning in separation logic. \emph{Logical Methods in Computer Science}. \href{https://iris-project.org/actris/}{Paper and mechanization}.

\leavevmode\protect\phantomsection\label{ref-aneris}{}{[}18{]} Aneris project. \emph{Program logic for developing and verifying distributed systems}. \href{https://github.com/logsem/aneris}{Development and documentation}.

\leavevmode\protect\phantomsection\label{ref-gondelman-rclib-2023}{}{[}19{]} Gondelman, L., Hinrichsen, J. K., Pereira, M., Timany, A., and Birkedal, L. (2023). Verifying reliable network components in a distributed separation logic with dependent separation protocols. \emph{Proceedings of the ACM on Programming Languages}, ICFP. \href{https://iris-project.org/pdfs/2023-icfp-relcom.pdf}{Paper}.

\leavevmode\protect\phantomsection\label{ref-benton-rhl-2004}{}{[}20{]} Benton, N. (2004). Simple relational correctness proofs for static analyses and program transformations. In \emph{Proceedings of POPL}, 14--25. \href{https://doi.org/10.1145/964001.964003}{DOI:10.1145/964001.964003}.

\leavevmode\protect\phantomsection\label{ref-li-foxtrot-2026}{}{[}21{]} Li, K. H., Aguirre, A., Tassarotti, J., and Birkedal, L. (2026). Contextual refinement of higher-order concurrent probabilistic programs. \emph{Proceedings of the ACM on Programming Languages}, 10(PLDI), Article 187. \href{https://doi.org/10.1145/3808265}{DOI:10.1145/3808265}. \href{https://clutch-project.org/pdfs/2026-pldi-foxtrot.pdf}{Paper}.

\leavevmode\protect\phantomsection\label{ref-timany-trillium-2024}{}{[}22{]} Timany, A., Gregersen, S. O., Stefanesco, L., Hinrichsen, J. K., Gondelman, L., Nieto, A., and Birkedal, L. (2024). Trillium: Higher-order concurrent and distributed separation logic for intensional refinement. \emph{Proceedings of the ACM on Programming Languages}, 8(POPL), Article 9. \href{https://iris-project.org/pdfs/2024-popl-trillium.pdf}{Paper}.

\leavevmode\protect\phantomsection\label{ref-gregersen-clutch-2024}{}{[}23{]} Gregersen, S. O., Aguirre, A., Haselwarter, P. G., Tassarotti, J., and Birkedal, L. (2024). Asynchronous probabilistic couplings in higher-order separation logic. \emph{Proceedings of the ACM on Programming Languages}, 8(POPL), Article 26. \href{https://doi.org/10.1145/3632868}{DOI:10.1145/3632868}. \href{https://clutch-project.org/pdfs/2024-popl-clutch.pdf}{Paper}.

\leavevmode\protect\phantomsection\label{ref-haselwarter-approxis-2025}{}{[}24{]} Haselwarter, P. G., Li, K. H., Aguirre, A., Gregersen, S. O., Tassarotti, J., and Birkedal, L. (2025). Approximate relational reasoning for higher-order probabilistic programs. \emph{Proceedings of the ACM on Programming Languages}, 9(POPL), Article 41. \href{https://doi.org/10.1145/3704877}{DOI:10.1145/3704877}. \href{https://clutch-project.org/pdfs/2025-popl-approxis.pdf}{Paper}.

\leavevmode\protect\phantomsection\label{ref-karwowski-geometry-2026}{}{[}25{]} Karwowski, J., Nielsen, F., and Staton, S. (2026). Compositional information geometry of probabilistic programs. \href{https://arxiv.org/abs/2610.17886v1}{arXiv:2610.17886v1}.

\leavevmode\protect\phantomsection\label{ref-haynes-fps-2026}{}{[}26{]} Haynes, H. (2026). Fixed-point scaffolding in the Clef programming language. \href{https://arxiv.org/abs/2606.02854}{arXiv:2606.02854}.

\leavevmode\protect\phantomsection\label{ref-islaris}{}{[}27{]} Islaris project. \emph{Verification of machine code against authoritative ISA semantics}. \href{https://github.com/rems-project/islaris}{Development and examples}.

\leavevmode\protect\phantomsection\label{ref-stroustrup-negative-overhead-2026}{}{[}28{]} Peterman, R. (2026, May 18). \href{https://www.developing.dev/p/creator-of-c-bell-labs-negative-overhead}{Creator of C++: Bell Labs, Negative Overhead Abstraction, Mistakes {|} Bjarne Stroustrup}. Interview with Bjarne Stroustrup. \emph{The Peterman Pod}. Transcript, 1:22:18--1:23:42.

\leavevmode\protect\phantomsection\label{ref-grunwald-vitanyi-2010}{}{[}29{]} Grünwald, P., and Vitányi, P. (2010). Shannon information and Kolmogorov complexity. Author manuscript dated July 22, 2010. \href{https://homepages.cwi.nl/~paulv/papers/info.pdf}{Manuscript}.

\leavevmode\protect\phantomsection\label{ref-solomonoff-1964}{}{[}30{]} Solomonoff, R. J. (1964). A formal theory of inductive inference. \emph{Information and Control}, 7(1):1--22.

\leavevmode\protect\phantomsection\label{ref-leike-hutter-2015}{}{[}31{]} Leike, J., and Hutter, M. (2015). On the computability of Solomonoff induction and knowledge-seeking. \href{https://hutter1.net/publ/sikscplex.pdf}{Author manuscript}.

\leavevmode\protect\phantomsection\label{ref-urschel-growth-2026}{}{[}32{]} Urschel, J. (2026). On the growth factor of random matrices. \href{https://arxiv.org/abs/2610.06785v1}{arXiv:2610.06785v1}.

\leavevmode\protect\phantomsection\label{ref-surace-metric-2020}{}{[}33{]} Surace, S. C., Pfister, J.-P., Gerstner, W., and Brea, J. (2020). On the choice of metric in gradient-based theories of brain function. \emph{PLOS Computational Biology}, 16(4):e1007640. \href{https://doi.org/10.1371/journal.pcbi.1007640}{Paper}.

\leavevmode\protect\phantomsection\label{ref-flugel-multitangent-2026}{}{[}34{]} Flügel, K., Coquelin, D., Weiel, M., Debus, C., Streit, A., and Götz, M. (2026). Beyond backpropagation: Optimization with multi-tangent forward gradients. \href{https://arxiv.org/abs/2410.17764}{arXiv:2410.17764}. Revised January 2026.

\leavevmode\protect\phantomsection\label{ref-bauschke-combettes-2011}{}{[}35{]} Bauschke, H. H., and Combettes, P. L. (2011). \emph{Convex Analysis and Monotone Operator Theory in Hilbert Spaces}. Springer.

\leavevmode\protect\phantomsection\label{ref-raissi-pinns-2019}{}{[}36{]} Raissi, M., Perdikaris, P., and Karniadakis, G. E. (2019). Physics-informed neural networks: A deep learning framework for solving forward and inverse problems involving nonlinear partial differential equations. \emph{Journal of Computational Physics}, 378:686--707.

\leavevmode\protect\phantomsection\label{ref-cheng-engram-2026}{}{[}37{]} Cheng, X., Zeng, W., Dai, D., et al. (2026). Conditional memory via scalable lookup: A new axis of sparsity for large language models. \href{https://arxiv.org/abs/2601.07372v1}{arXiv:2601.07372v1}.

\leavevmode\protect\phantomsection\label{ref-gordon-lcf-hol-2000}{}{[}38{]} Gordon, M. J. C. (2000). From LCF to HOL: A short history. In G. Plotkin, C. Stirling, and M. Tofte (eds.), \emph{Proof, Language, and Interaction}. MIT Press. \href{https://www.cl.cam.ac.uk/archive/mjcg/papers/HolHistory.pdf}{Author manuscript}.

\leavevmode\protect\phantomsection\label{ref-petricek-coeffects-2014}{}{[}39{]} Petricek, T., Orchard, D., and Mycroft, A. (2014). Coeffects: A calculus of context-dependent computation. In \emph{Proceedings of ICFP}.

\leavevmode\protect\phantomsection\label{ref-armstrong-erlang-2003}{}{[}40{]} Armstrong, J. (2003). \emph{Making reliable distributed systems in the presence of software errors}. Doctoral dissertation, Royal Institute of Technology, Stockholm. \href{https://erlang.org/download/armstrong_thesis_2003.pdf}{Thesis}.

\leavevmode\protect\phantomsection\label{ref-wilcox-verdi-2015}{}{[}41{]} Wilcox, J. R., Woos, D., Panchekha, P., Tatlock, Z., Wang, X., Ernst, M. D., and Anderson, T. (2015). Verdi: A framework for implementing and formally verifying distributed systems. \emph{PLDI}. \href{https://homes.cs.washington.edu/~ztatlock/pubs/verdi-wilcox-pldi15.pdf}{Paper}.

\leavevmode\protect\phantomsection\label{ref-refinedc}{}{[}42{]} RefinedC project. \emph{Automating the foundational verification of C code with refined ownership types}. \href{https://plv.mpi-sws.org/refinedc/}{Paper and artifact}.

\leavevmode\protect\phantomsection\label{ref-castro-mcerlang-2011}{}{[}43{]} Castro, D., Gulías, V. M., Benac Earle, C., Fredlund, L.-Å., and Rivas, S. (2011). A case study on verifying a supervisor component using McErlang. \emph{Electronic Notes in Theoretical Computer Science}, 271, 23--40. \href{https://dcastrop.github.io/files/2011-prole-mcerlang.pdf}{Paper}.

\leavevmode\protect\phantomsection\label{ref-gavranovic-categorical-2024}{}{[}44{]} Gavranovic, B., Lessard, P., Dudzik, A., et al. (2024). Categorical deep learning is an algebraic theory of all architectures. \href{https://arxiv.org/abs/2402.15332}{arXiv:2402.15332}.

\leavevmode\protect\phantomsection\label{ref-tarski-1951}{}{[}45{]} Tarski, A. (1951). \emph{A Decision Method for Elementary Algebra and Geometry}, 2nd edition. University of California Press.

\leavevmode\protect\phantomsection\label{ref-matiyasevich-1970}{}{[}46{]} Matiyasevich, Y. V. (1970). Enumerable sets are Diophantine. \emph{Doklady Akademii Nauk SSSR}, 191:279--282.
\end{document}